\documentclass[%
 reprint,
 amsmath,amssymb,
 aps,
nofootinbib,
superscriptaddress
]{revtex4-1}

\usepackage{graphicx}
\usepackage{dcolumn}
\usepackage{bm}
\usepackage{enumitem}
\usepackage[draft, inline, nomargin]{fixme}
\fxsetup{theme=color, mode=multiuser}
\FXRegisterAuthor{fs}{2}{\color{blue}FS}
\usepackage{lineno}

\begin{document}


\title{Measurement of Muon-induced High-energy Neutrons from \\ Rock in an Underground Gd-doped Water Detector}

\author{F. Sutanto}
 \email{fsutanto@umich.edu}
 \affiliation{Department of Nuclear Engineering and Radiological Sciences, University of Michigan, Ann Arbor, MI 48109, USA}
 \affiliation{Lawrence Livermore National Laboratory, Livermore, CA 94550, USA}
\author{O.A. Akindele}
 \affiliation{Lawrence Livermore National Laboratory, Livermore, CA 94550, USA}
\author{M. Askins}
 \affiliation{Physics Department, University of California, Davis, CA 95616, USA}
  \email{Present address: Physics Department, University of California, Berkeley, CA 94720, USA}
\author{M. Bergevin}
 \email{bergevin1@llnl.gov}
 \affiliation{Lawrence Livermore National Laboratory, Livermore, CA 94550, USA}
\author{A. Bernstein}
 \affiliation{Lawrence Livermore National Laboratory, Livermore, CA 94550, USA}
\author{N.S. Bowden}
 \affiliation{Lawrence Livermore National Laboratory, Livermore, CA 94550, USA}
\author{S. Dazeley}
 \email{dazeley2@llnl.gov}
 \affiliation{Lawrence Livermore National Laboratory, Livermore, CA 94550, USA}
\author{P. Jaffke}
 \affiliation{Center for Neutrino Physics, Virginia Tech, Blacksburg, VA 24061, USA}
\author{I. Jovanovic}
 \affiliation{Department of Nuclear Engineering and Radiological Sciences, University of Michigan, Ann Arbor, MI 48109, USA}
\author{S. Quillin}
 \affiliation{Atomic Weapons Establishment, Aldermaston, Reading RG7 4PR, United Kingdom}
\author{C. Roecker}
 \affiliation{Department of Nuclear Engineering, University of California, Berkeley, CA 94720, USA}
 \email{Present address: Los Alamos National Laboratory, Los Alamos, NM 87545, USA}
\author{S.D. Rountree}
 \affiliation{Center for Neutrino Physics, Virginia Tech, Blacksburg, VA 24061, USA}

\date{\today}

\begin{abstract}
\noindent We present a measurement of the rate of correlated neutron captures in the WATCHBOY detector, deployed at a depth of approximately 390 meters water equivalent (m.w.e.) in the Kimballton Underground Research Facility (KURF). WATCHBOY consists of a cylindrical 2 ton water target doped with 0.1\% gadolinium, surrounded by a 40 ton undoped water hermetic shield. We present a comparison of our results with the expected rate of correlated neutron captures arising from high-energy neutrons incident on the outside of the WATCHBOY shield, predicted by a hybrid FLUKA/GEANT4-based simulation. The incident neutron energy distribution used in the simulation was measured by a fast neutron spectrometer, the 1.8-ton Multiplicity and Recoil Spectrometer (MARS) detector, at the same depth. We find that the measured detection rate of two correlated neutrons is consistent with that predicted by simulation. The result lends additional confidence in the detection technique used by MARS, and therefore in the MARS spectra as measured at three different depths. Confirmation of the fast neutron flux and spectrum is important as it helps validate the scaling models used to predict the fast neutron fluxes at different overburdens.

\end{abstract}

\pacs{Valid PACS appear here}
\maketitle

\section{\label{sec:level1}Introduction}

Muon--induced fast neutrons are an important background for many rare event measurements in underground laboratories. 
Fast neutrons were a possible source of the discrepancy between the CDMS~\cite{abusaidi} and the DAMA/LIBRA dark matter experiments~\cite{bernabei,klinger,barbeau}. High-energy neutrons were also shown to be the dominant background in the relatively shallow reactor neutrino oscillation experiment 
such as Double Chooz~\cite{abe,abe2}, Daya Bay~\cite{dayaBay}, and RENO~\cite{reno2015}. 
Despite their importance, it is difficult to unambiguously measure the neutron flux and energy distribution at depth, because the flux is low and generally rare event detectors are designed to moderate and shield themselves from the fast neutron flux rather than measure it.

In order to plan and design experiments at a range of depths, existing models used to scale high-energy neutron flux in rock must be better validated than current data allows~\cite{Bernstein}. Motivating the present work, a kiloton scale gadolinium-doped light water Cherenkov detector WATCHMAN (WATer CHerenkov Monitor of ANtineutrinos) is planned in the near future in order to demonstrate remote monitoring of the antineutrino flux arising from the Hartlepool reactor in Northern England~\cite{askin}. In such detectors, the antineutrino signal appears as two interactions (a positron followed by a neutron capture) correlated in time and position~\cite{giunti}. Backgrounds include accidental coincidences of uncorrelated single events, antineutrinos from other reactors, and correlated signals arising from muon induced radionuclides and fast neutrons. Two experiments, MARS~\cite{roecker} and WATCHBOY~\cite{dazeley}, were built to measure fast neutrons and long-lived radionuclides respectively, with the goal of improving the predictive capability for large detectors such as WATCHMAN.  

The MARS detector was deployed at depths of 390 m.w.e., 540 m.w.e., and 1450 m.w.e. within the Kimballton mine near the Kimballton Underground Research Facility (KURF) in Virginia. The purpose of the deployment was to provide a set of neutron spectra measurements at different depths, measured with the same detector and systematic effects. 
MARS neutron spectra were published elsewhere, along with a detailed description of the detector~\cite{RoeckerThesis}.
Briefly, MARS was a transportable fast neutron detection system that consisted of two plastic scintillator-based neutron detectors mounted above and below a lead spallation target. 
The dimensions of the lead target were 101~cm long by 71~cm wide by 20~cm high (1.8 tons). 
The dimensions of both neutron detectors were similar -- 100~cm$\times$75~cm$\times$25~cm.
Each neutron detector consisted of twelve 2~cm thick plastic scintillator layers interleaved with gadolinium coated Mylar sheets.
The detection method was as follows: high-energy neutrons initiate a spallation reaction in the lead target and generate secondary neutrons. The number of secondary neutrons produced is correlated with the energy of the initiating fast neutron. A significant fraction of these secondary neutrons down-scatter in the scintillator layer, thermalize and eventually capture on a gadolinium nucleus an average of 18.7$\pm$3.0~$\mu s$ later~\cite{roecker}. The de-excitation of the gadolinium nucleus upon capture generates several gamma-rays with total energy of 7.9~MeV and 8.5~MeV for $^{157}$Gd and $^{155}$Gd, respectively. These gamma-rays interact and partially deposit their energy in the plastic scintillator. 

The primary aim of WATCHBOY was to measure the rate of radionuclide production in water due to the passage of high energy cosmic ray muons through its 2 ton water target~\cite{dazeley}. The target was surrounded by a hermetic 40 ton undoped water muon detector and neutron shield, the purpose of which was to stop high energy neutrons from reaching the target, a potential source of backgrounds for the radionuclide measurement. Nevertheless, a small fraction of the highest energy neutrons were able to penetrate the shield and make it to the target, producing correlated groups of neutron captures via spallation in the target volume. Since WATCHBOY was a large and immobile detector, it was deployed at a single depth, which was estimated to be 390 m.w.e. - the same depth as one of the MARS deployments.

WATCHBOY was conceived as a prototype for WATCHMAN.
The radionuclide considered to be of greatest importance to WATCHMAN was $^{9}$Li ($\tau$ = 257 ms, Q value = 11.9 MeV), which decays via the simultaneous emission of a multi-MeV beta particle and a neutron, similar to the antineutrino IBD reaction. While WATCHBOY was intended for measuring radionuclides, its dominant background was caused by energetic neutrons produced in the surrounding rock by muons passing near the detector. These neutrons can enter the WATCHBOY target undetected by the outer shield, and initiate spallation reactions and nuclear recoils in the detector~\cite{galbiati} which lead to correlated neutron capture signals in the WATCHBOY target. 

Also of great interest to WATCHMAN was the development of a simulation code capable of predicting the rate, via spallation, of free neutrons in water-based detectors at any depth. To address this need, this work has two aims: first, to measure the rate of correlated neutron capture candidates in the (Gd-water) WATCHBOY target, and second, to determine the accuracy of two candidate simulation codes (GEANT4 and FLUKA) at predicting the observed correlated event rates using the independent MARS measurement of the neutron energy spectrum as input.

The WATCHBOY detector and its data acquisition system (DAQ) are described in Section~\ref{sec:level2}. Data selection and analysis are described in Section~\ref{sec:level3}. The simulation model is presented in Section~\ref{sec:leve}. Section~\ref{sec:level4} describes our measurements of the multiplicity of correlated neutron-like events. Finally, conclusions are presented in Section~\ref{sec:level5}.

\vspace{-0.5cm}
\section{\label{sec:level2}Watchboy detector}

\begin{figure}[!ht]
\centering
\begin{minipage}{0.50\textwidth}
\includegraphics[width=\linewidth]{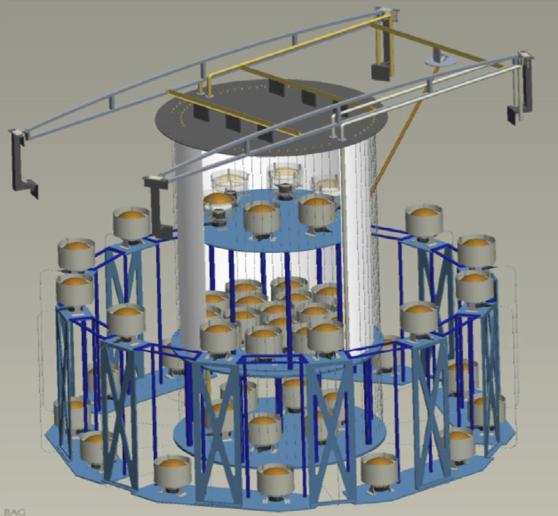}
\end{minipage}
\caption{The PMT arrangement and supporting structure inside the WATCHBOY detector~\cite{dazeley}. The containment bag was filled with Gd-doped water and consisted of two optically separated regions-- a lower 2-ton target region instrumented with 16~PMTs and an upper veto region instrumented with 6~PMTs. The containment bag was surrounded by 40-ton pure water veto volume containing 30~PMTs. WATCHBOY was deployed at a depth of $\approx$390~m.w.e. within KURF.}\label{Fig:watchboySetup}
\end{figure}

Figure ~\ref{Fig:watchboySetup} shows a schematic of the detector components inside WATCHBOY, with the outer tank walls not displayed~\cite{dazeley}. A cutaway view of the gadolinium doped inner containment bag is shown in the center. The  containment bag was filled with Gd-doped water (0.2\% GdCl$_3$) and consisted of two optically separated volumes - a lower 2~ton ``target'' region, equipped with 16--tightly packed upward facing 10'' Hamamatsu R7081 photomultiplier tubes (PMTs), and an upper region which formed part of the muon veto. 
Gadolinium was chosen for its high neutron absorption cross section (49,000 barns for natural Gd). 
Cherenkov light produced following neutron capture on Gd in the target volume was collected on the 16 upward-facing PMTs. The target walls were coated with a highly diffusely reflective Teflon to enhance light collection.
Due to the reflectivity of the walls and the small number of PMTs, most of the detected light followed multiple reflections. The effect was to distribute ($\approx$9~ns time spread) the light fairly evenly among the target PMTs whenever Cherenkov light was produced by a physics event. 
As mentioned earlier, above the target and within the same containment bag was an optically separated top volume which formed part of the muon veto equipped with 6 upward facing 10'' Hamamatsu R7081 PMTs. Since it existed within the same containment bag, it was also filled with Gd-doped water.
Surrounding the bag was a $\approx$40 ton pure water volume containing 30 additional 10'' PMTs that also formed part of the muon veto. 

Four 16-channel 250-MHz 14-bit SIS3316 Struck ADC/digitizer boards were used to collect the PMT signals. PMT signals were summed into groups of four to create a trigger. Triggers were formed whenever any summed group reached a threshold of $\approx$1~photoelectron (PE).
If any of the discriminators were triggered, the pulse integral of all 52 channels were then read-out to disk.

\section{\label{sec:level3}Data selection and analysis}

Ad hoc neutron selection criteria were developed and discussed in detail in the previous WATCHBOY paper~\cite{dazeley}.
Here we discuss only the pertinent highlights. 

One of the most prevalent backgrounds in WATCHBOY was caused by the PMTs---flashers and other noise triggers, which predominantly produced signals within only a single PMT. A straightforward way to remove these noise triggers was to define a parameter sensitive to the ``evenness'' of the distribution of light among the PMTs - called ``charge balance ($Q_b$).'' This parameter was defined as follows:

\begin{equation} \label{Eq:twelve}
Q_b = \sqrt{\frac{\sum{Q_i^2}}{Q_{sum}^2} - \frac{1}{N} }
\end{equation}

\noindent where $N$ is the total number of target PMTs, $Q_i$ is the charge collected by the $i$-th PMT, and $Q_{sum}$ is the summed charge of all 16 target PMTs. A charge balance value close to zero indicates evenly distributed signal among all 16 target PMTs, and close to one indicates the opposite extreme (most of the light detected was by a single PMT).

\begin{figure}[!ht]
\centering
\begin{minipage}{0.5\textwidth}
\includegraphics[width=\linewidth]{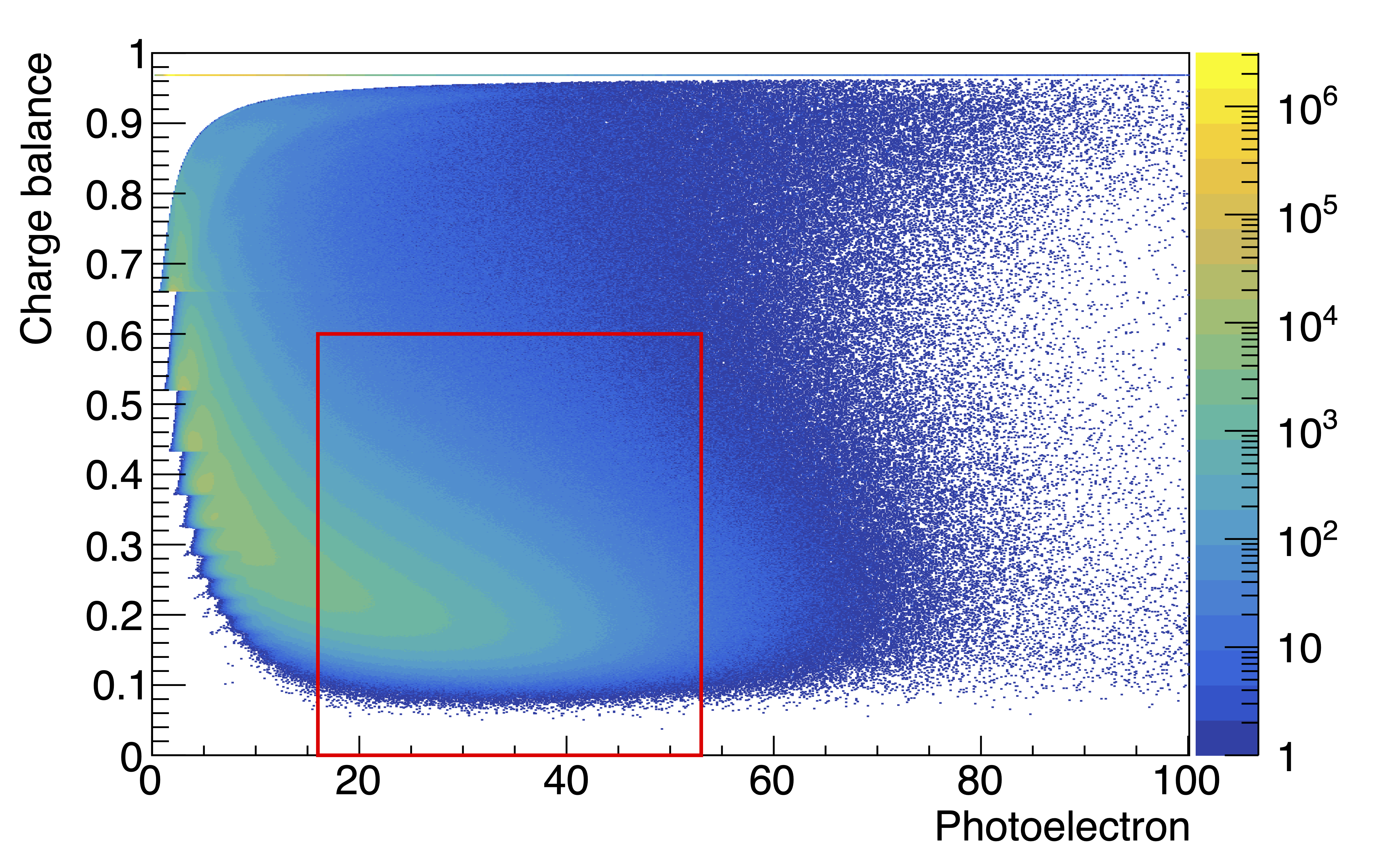}
\vspace{-0.4cm}
\end{minipage}
\caption{The detector response to a $^{252}$Cf source in terms of the total detected photolectrons and the evenness of the light distribution as defined in Eq.(\ref{Eq:twelve}). The analysis cuts that select for neutron capture candidate~\cite{dazeley} are shown in red.}\label{Fig:qbpeCf}
\end{figure}

\begin{figure}[!ht]
\centering
\begin{minipage}{0.5\textwidth}
\includegraphics[width=\linewidth]{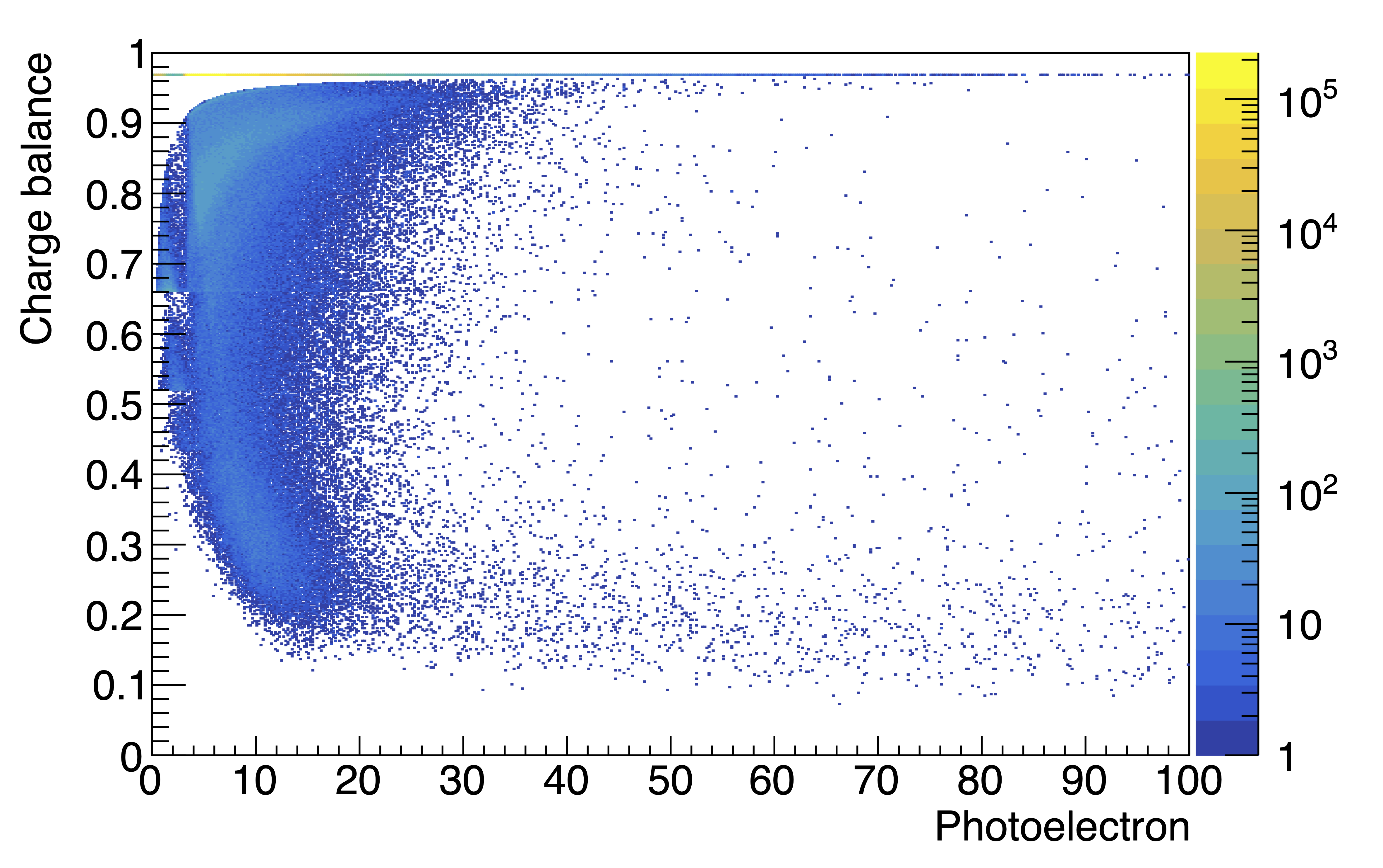}
\vspace{-0.4cm}
\end{minipage}
\caption{The detector response to background in terms of the total detected photolectrons and the evenness of the light distribution as defined in Eq.(\ref{Eq:twelve}).}\label{Fig:qbpePhysics}
\end{figure}

\begin{figure}[!ht]
\centering
\begin{minipage}{0.5\textwidth}
\includegraphics[width=\linewidth]{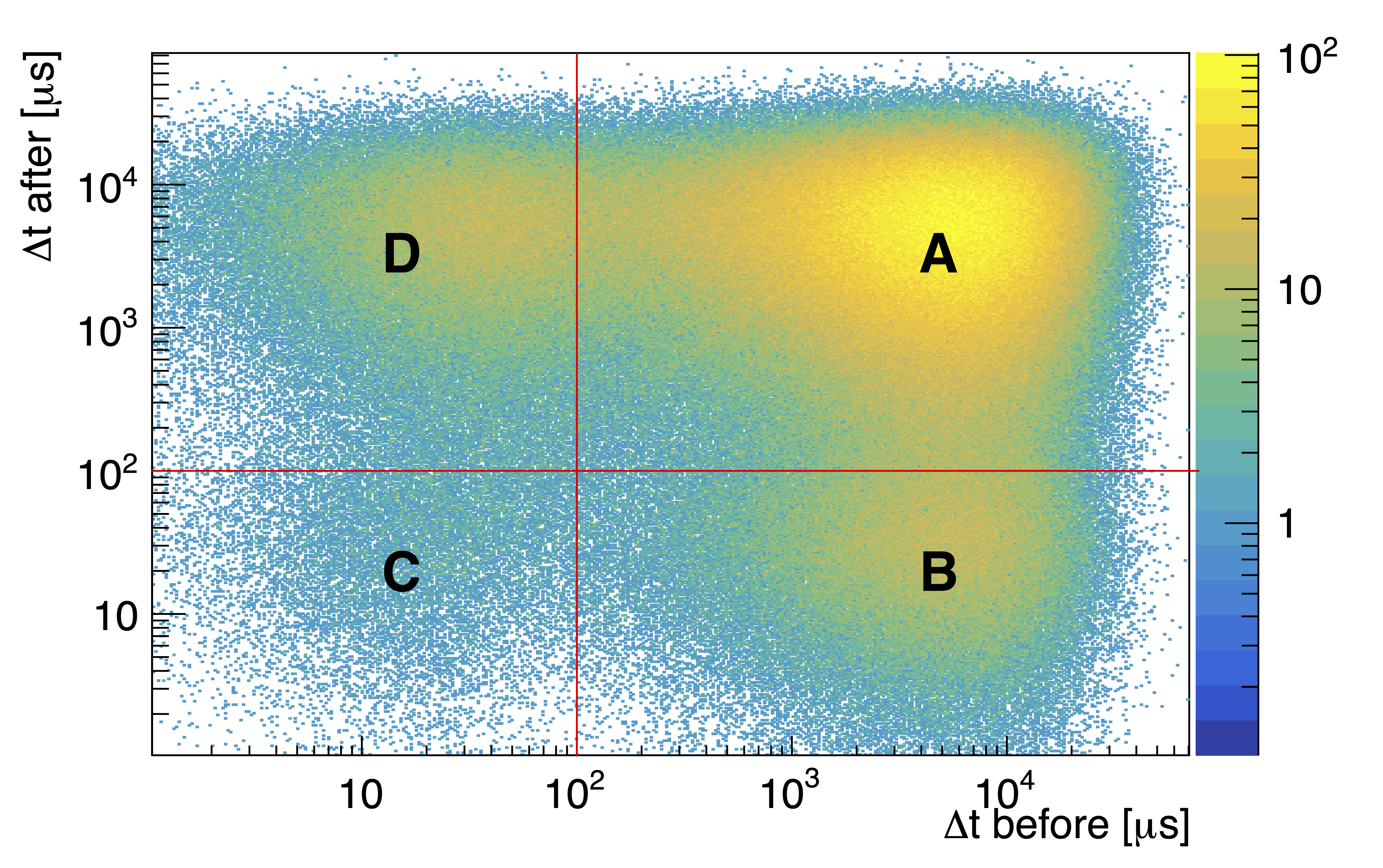}
\vspace{-0.4cm}
\end{minipage}
\caption{The distribution of time intervals between detected events. The $100$~$\mu$s temporal cut is shown in red. Region B, C, and D likely correspond to the correlated neutron events and region A corresponds to single events.}\label{Fig:t2mu}
\end{figure}

During calibrations, a $^{252}$Cf source was inserted through the yellow calibration tube, as shown in Fig.~\ref{Fig:watchboySetup}, and placed approximately 4 cm from the side wall of the inner detector. Figure~\ref{Fig:qbpeCf} shows charge balance as a function of detected photoelectrons for the $^{252}$Cf calibration data. From the calibration runs, neutron capture candidates were found to preferentially have charge-balance values less than 0.6. The summed number of photoelectrons were also larger on average than for background data (Fig.~\ref{Fig:qbpeCf} and Fig.~\ref{Fig:qbpePhysics}). The background data consisted primarily of the PMT flasher-based events described earlier, and also events caused by low energy gamma-rays from the PMT glass, which is known to contain potassium, thorium and uranium. Using the same neutron selection cuts as previously described in~\cite{dazeley}, we required the summed charge of the target PMTs to be in the range 16 to 53 photoelectrons. The upper limit helps to reject Michel electrons and muons that clip the target. 
To minimize background cosmogenic neutrons caused by muons traversing the detector, all events within 1 ms of a muon candidate event were rejected~\cite{dazeley}. Muon events were defined as any event with summed charge in the veto greater than 30~photoelectrons or summed charge in the target region greater than 100~photoelectrons.

Using $^{252}$Cf data, we isolated a subset of events highly likely to be neutron captures by selecting events correlated with a previous event~\cite{labranche}. The technique takes advantage of the fact that multiple neutrons are generally emitted simultaneously in $^{252}$Cf spontaneous fission events (3.76~prompt~neutrons/fission on average ~\cite{cf252Mul_1,cf252Mul_2}). Fig.~\ref{Fig:t2mu} shows a two dimensional picture of the inter-event time distribution observed in $^{252}$Cf calibration data, as explained below:
\begin{itemize}
\item Uncorrelated events appear in region A. These events have a long time interval to the prior event ($\Delta t$ \textit{before}) and a long time interval to the next event ($\Delta t$ \textit{after}).
\item The first event of a correlated group appears in region B; these events have a long time interval from the prior event and a short time interval to the next event. Many of these are likely to be caused by either fission gamma rays or neutron captures resulting from a fission event.
\item The last detected event from a correlated group appears in region D; with a short time interval from a prior event and a long time interval to the next event. This subset are preferentially populated by neutron captures following thermalization.
\item Correlated events with a short time interval to both the previous and next event are in region C; This subset is also preferentially populated by neutron captures following thermalization.
\end{itemize}
Notice that there must be at least two events in a correlated group to occupy region B and D. There must be at least three events in a correlated group to occupy region C. Region B corresponds to the first event, which could be either a prompt gamma-ray event or a delayed neutron event. Region C and region D always correspond to delayed events that are likely due to thermalized neutrons capturing in the water volume.
For a $^{252}$Cf calibration run, the rate of events observed in region A, B, C, and D were 252~Hz, 100~Hz, 30~Hz, and 100~Hz, respectively. The accidental probability of two correlated events within 100~$\mu s$ time window due to the single events can be approximated as (252~Hz)$^2\times$~100~$\mu s$ = 6.4~Hz. This implies that accidentals contribute to $\approx$5\% of the total events in region C and D.
Hence, the events that appear in region C and D are heavily dominated by neutron captures.

\section{\label{sec:leve} GEANT4 simulation }

\begin{figure}[ht!]
	\begin{center}
    \includegraphics[width=0.5\textwidth]{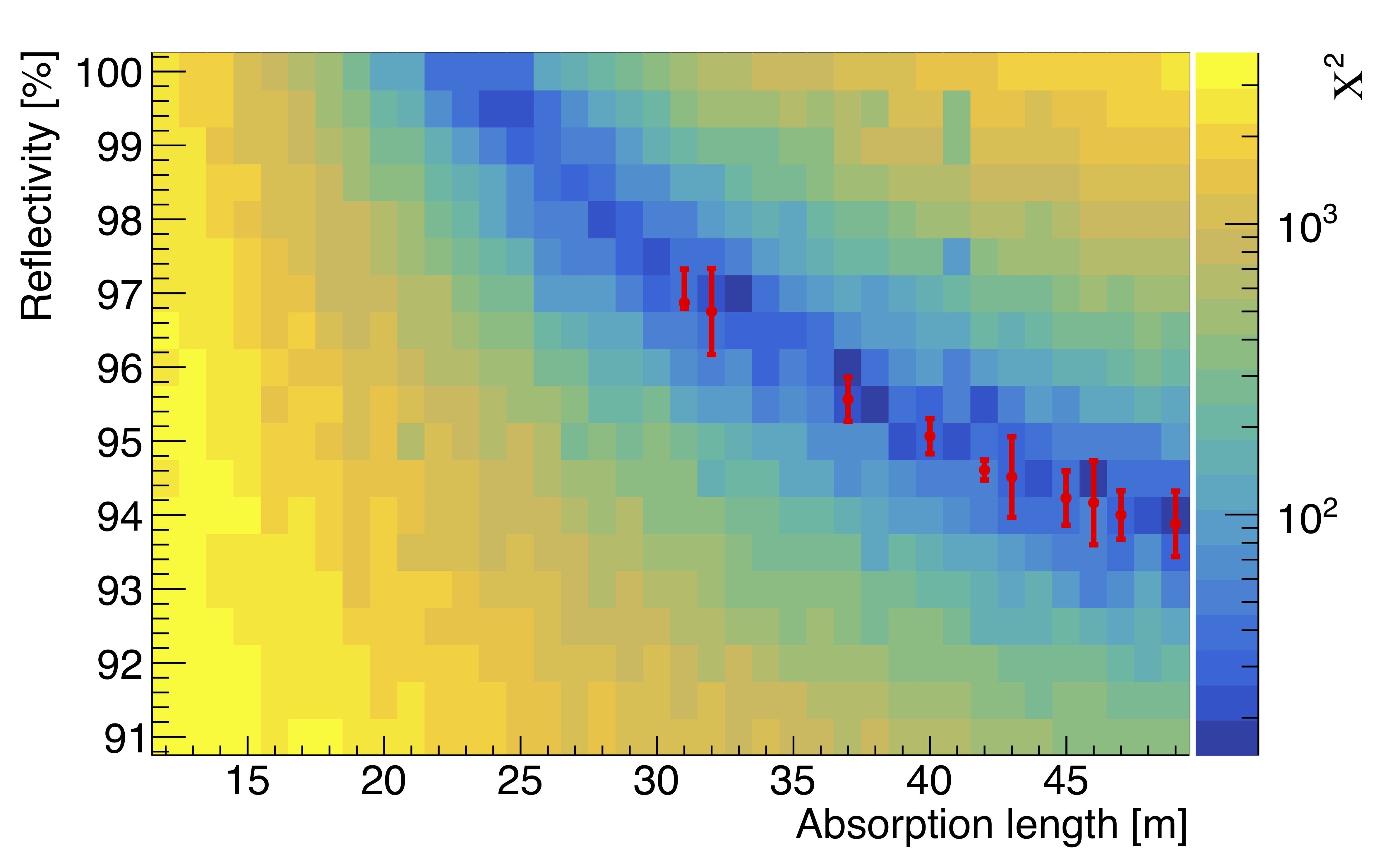}\\
    (a) \\
    \includegraphics[width=0.5\textwidth]{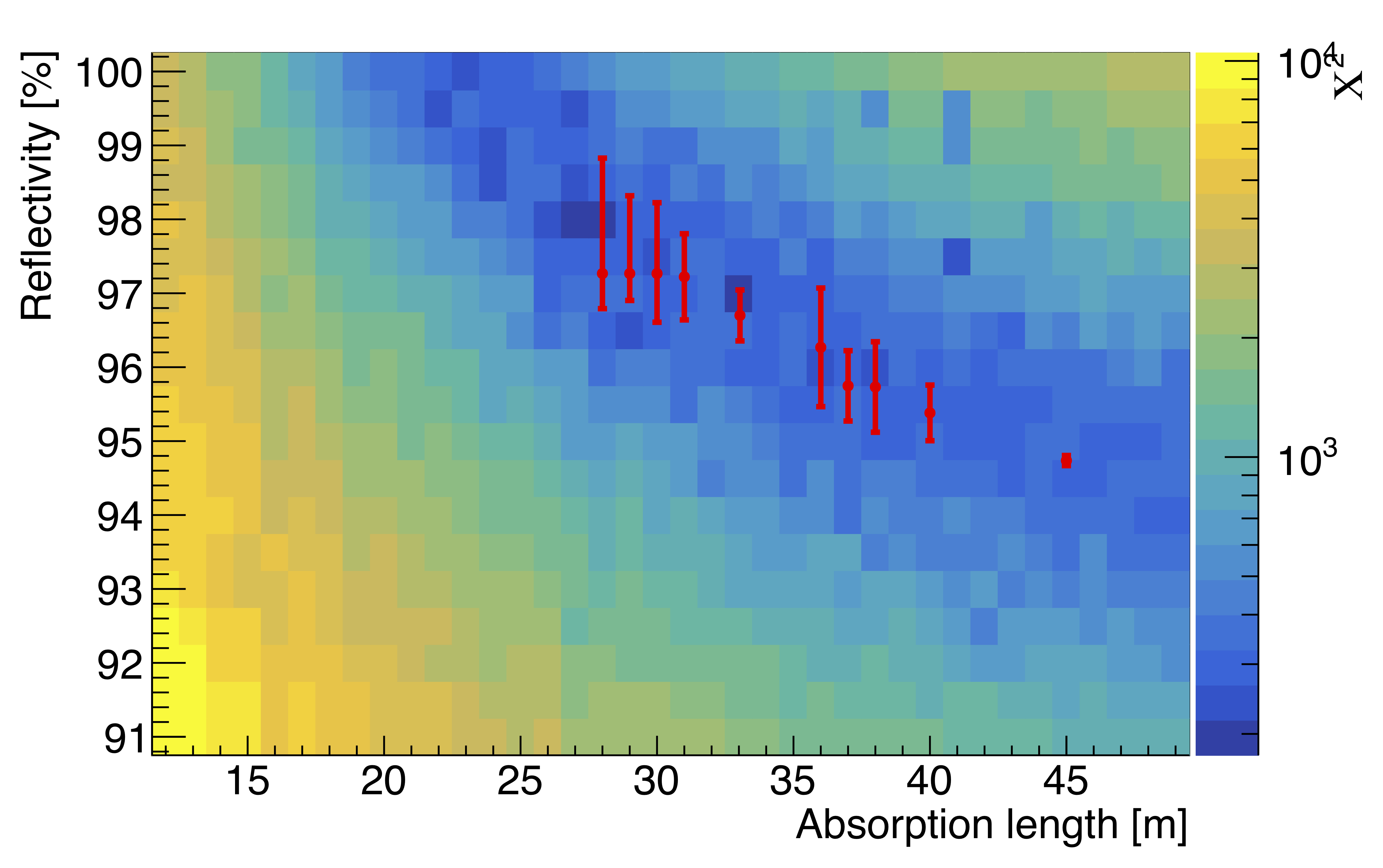}\\
    (b)
	\end{center}
    \caption{(Color online) The distribution of the goodness of fit ($\chi^2$/~ndf) is obtained by comparing (a) the simulated and $^{252}$Cf neutron capture candidate photoelectron distribution and (b) the similar comparison for the charge balance distributions. The lowest $\chi^2$ values are 15 (26 DOF) and 177 (59 DOF) for the (a) photoelectron and (b) charge balance comparisons respectively.}\label{Fig:chi}
\end{figure}

\begin{figure}[ht!]
	\begin{center}
    \includegraphics[width=0.5\textwidth]{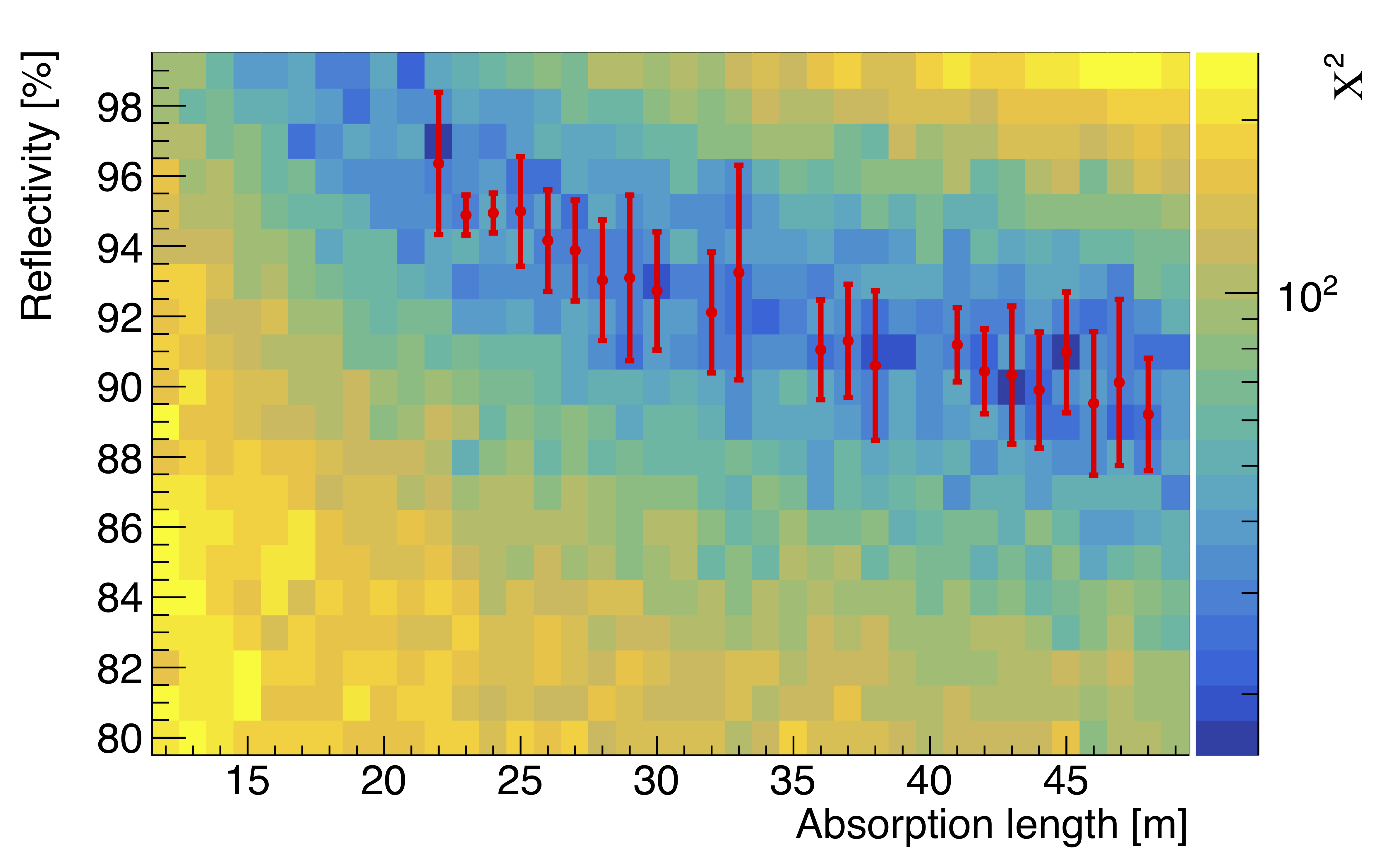}\\
    (a) \\
    \includegraphics[width=0.5\textwidth]{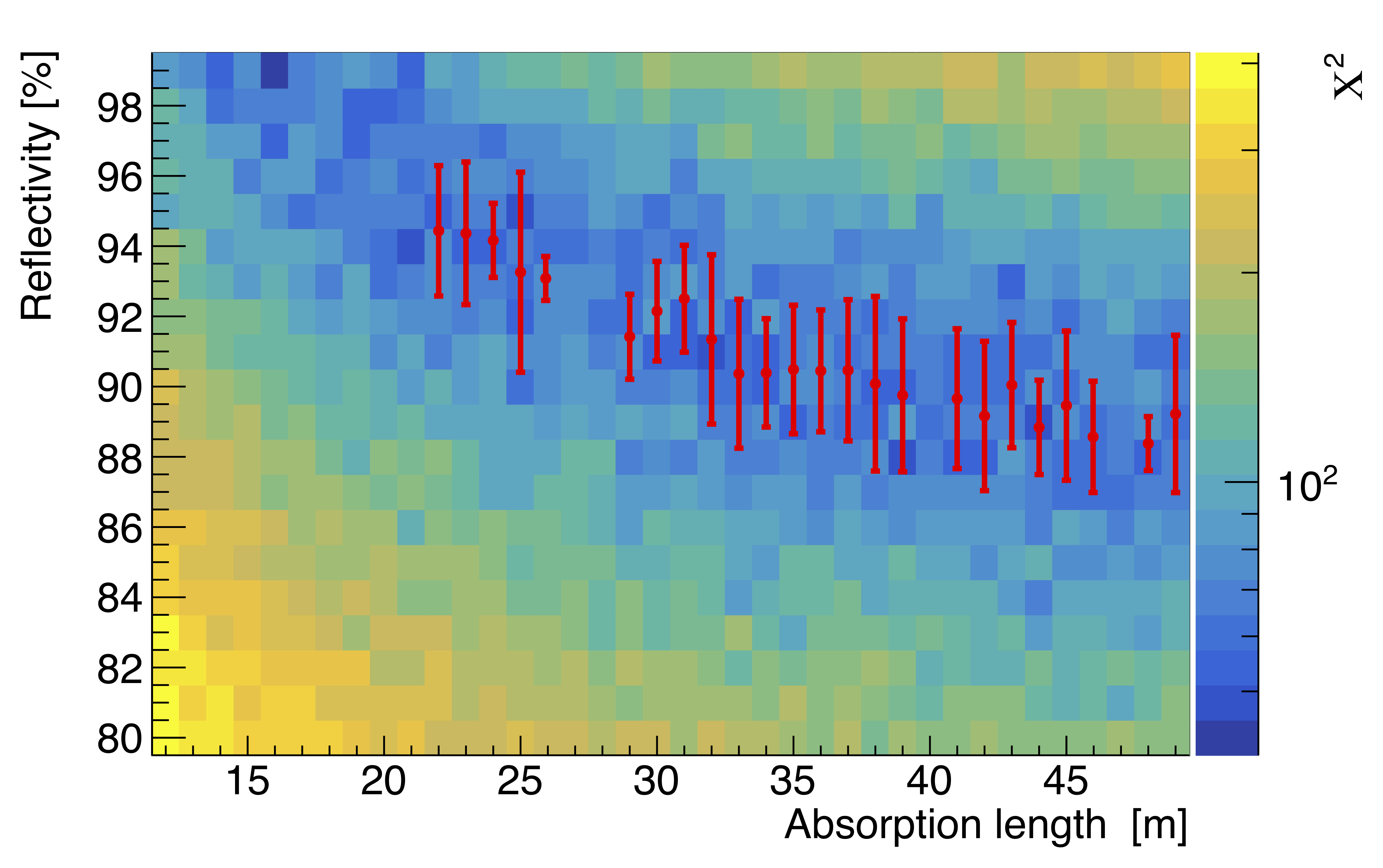}\\
    (b)
	\end{center}
    \caption{(Color online) The distribution of the goodness of fit ($\chi^2$/~ndf) is obtained by comparing (a) the simulated and muon-induced neutron candidate photoelectron distribution and (b) the similar comparison for the charge balance distributions. The lowest $\chi^2$ values are 16 (26 DOF) and 40 (51 DOF) for the (a) photoelectron and (b) charge balance comparisons respectively.}\label{Fig:chiMuonNeu}
\end{figure}

The GEANT4-based simulation framework RAT-PAC (Reactor Analysis Tool - Plus Additional Codes) was used to simulate neutron and gamma-ray interactions in the WATCHBOY detector and to determine the neutron capture efficiency. RAT-PAC was developed partly for KamLAND\footnote{ RAT incorporates parts of the scintillation and PMT simulation from Generic Liquid Scintillator GEANT4 simulation, written and maintained by Glenn Horton-Smith from the KamLAND collaboration.}, and partly for the Braidwood detector proposal~\cite{Bolton}, and has since been used and validated by the SNO+ collaboration. For this work, GEANT4 version 4.10.3 with the Shielding physics list and Photon Evaporation model were used. 

\begin{figure}[ht!]
	\begin{center}
    \includegraphics[width=0.5\textwidth]{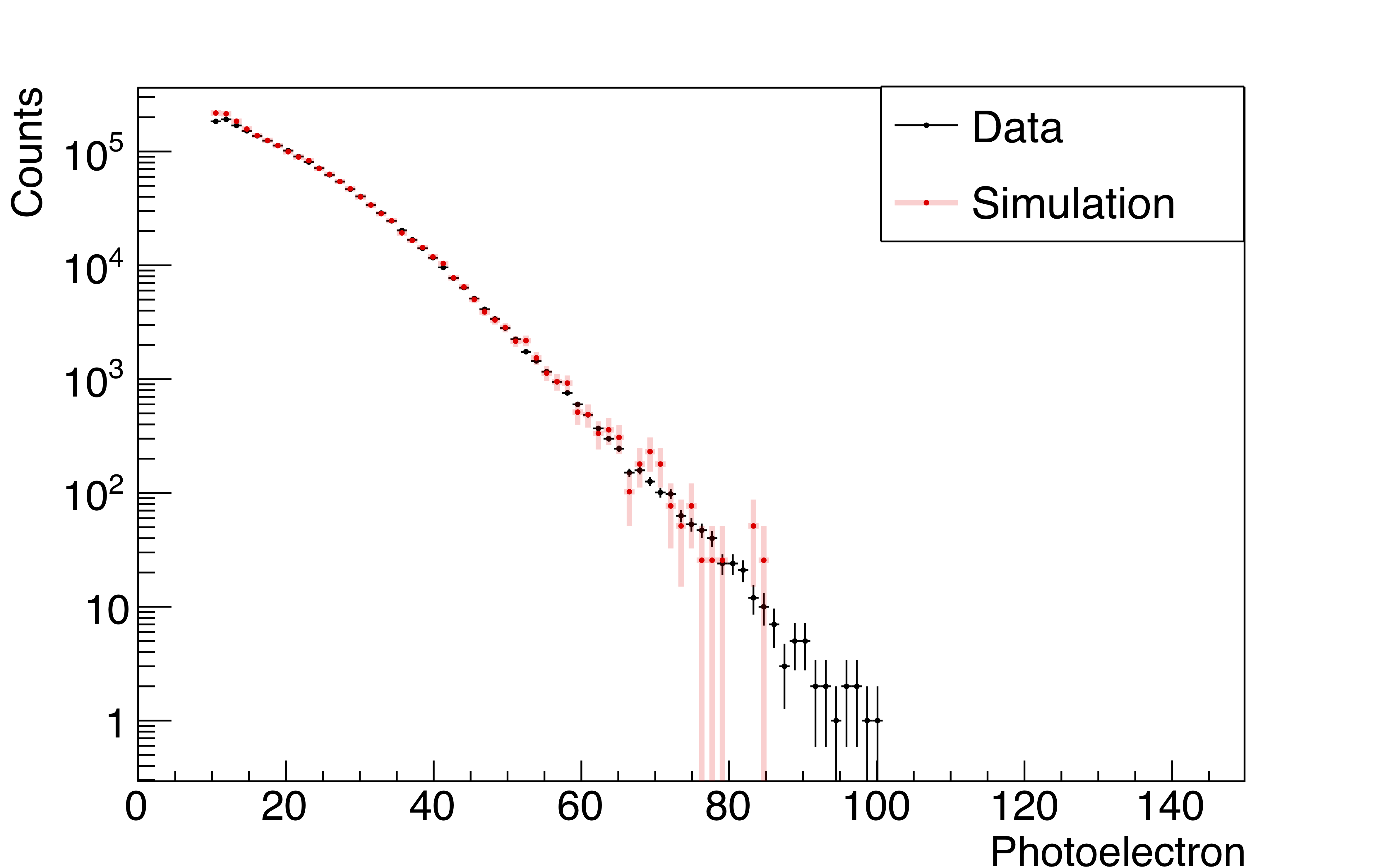}\\
    (a)\\
    \includegraphics[width=0.5\textwidth]{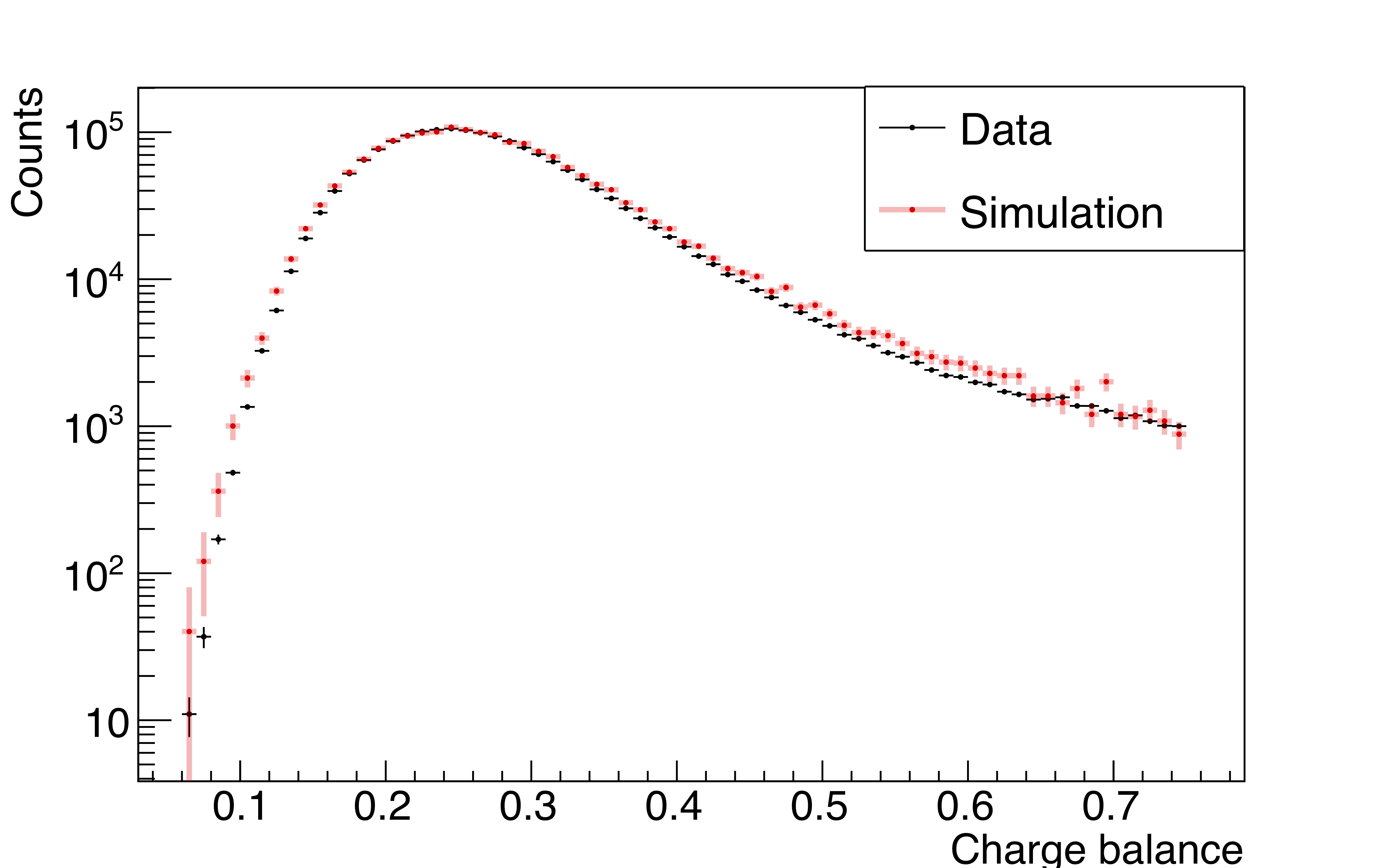}\\
    (b)
	\end{center}
    \caption{(Color online) Example of the comparison between (a) the simulated light collection and (b) the simulated charge balance responses to the WATCHBOY $^{252}$Cf calibration data.}\label{Fig:cf}
\end{figure}

\begin{figure}[ht!]
	\begin{center}
    \includegraphics[width=0.5\textwidth]{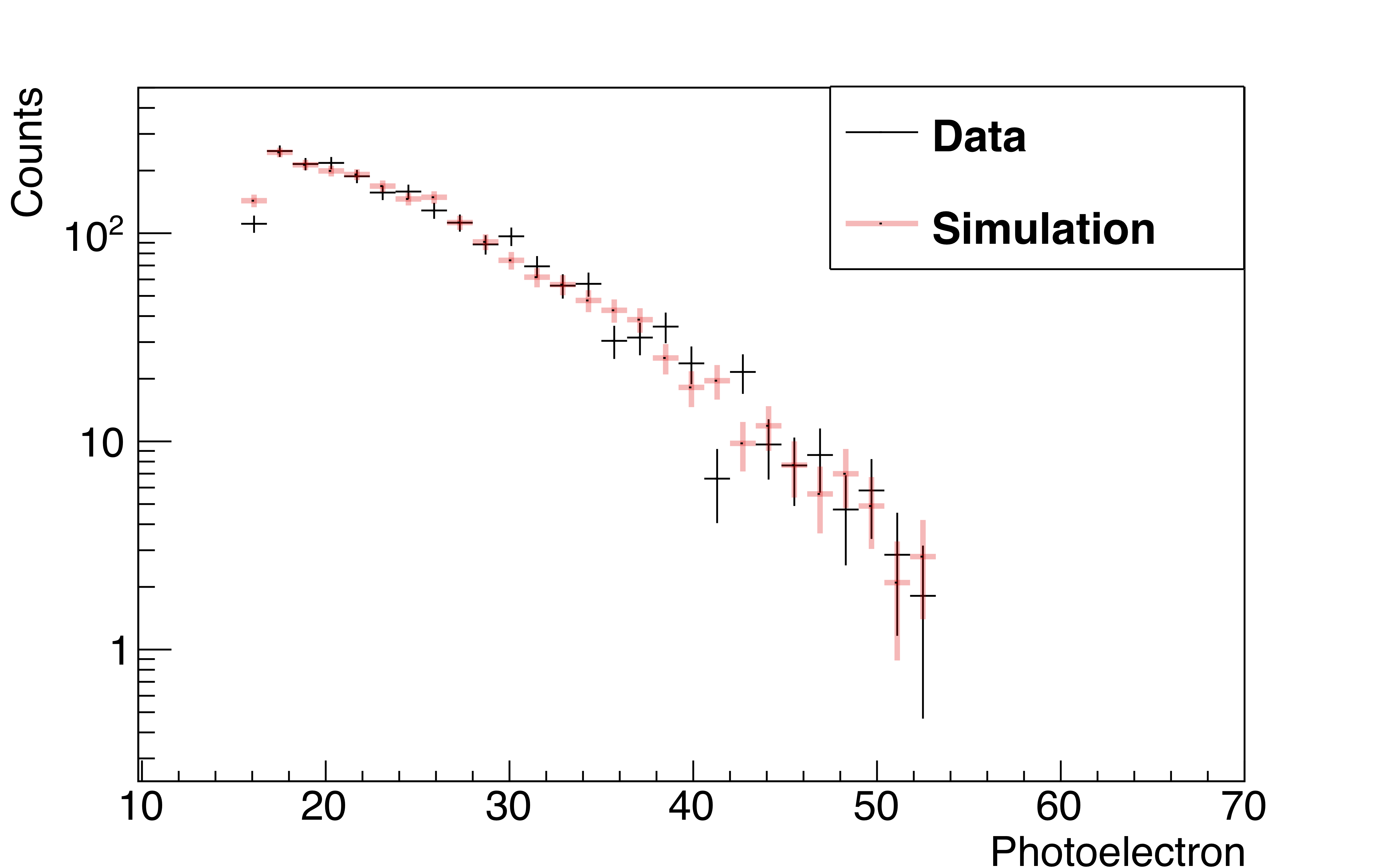}\\
    (a)\\
    \includegraphics[width=0.5\textwidth]{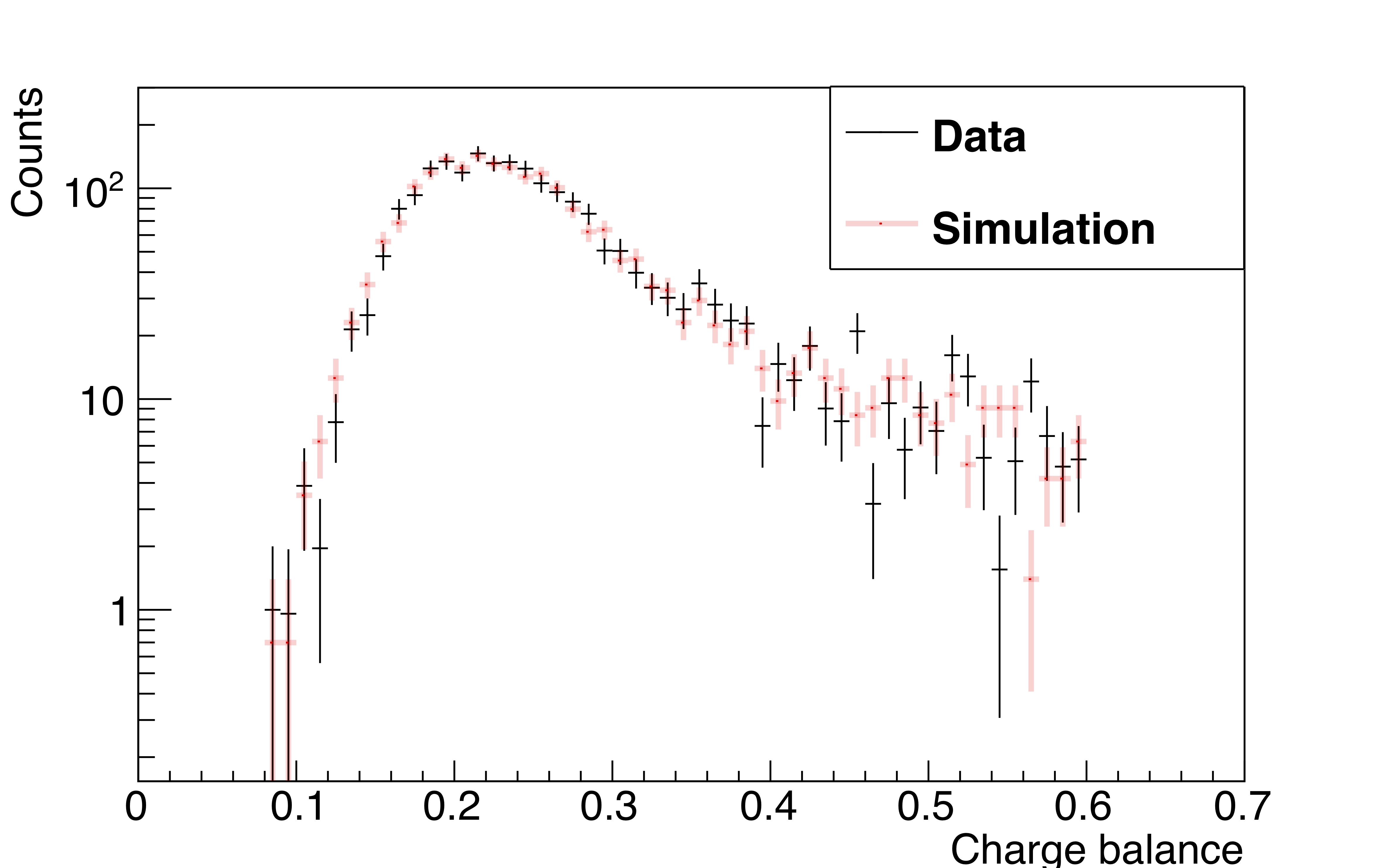}\\
    (b)
	\end{center}
    \caption{(Color online) Example of the comparison between (a) the simulated light collection and (b) the simulated charge balance responses to the WATCHBOY neutron data from muon spallation in the water target.}\label{Fig:muonNeu}
\end{figure}

In recent years, large discrepancies have been observed between GEANT4 and measured gamma-ray spectra and multiplicity distributions for neutron captures on gadolinium~\cite{yano,chen,hagiwaraGd}. DICEBOX is an alternative gamma-ray cascade model that appears to model gamma cascades of excited nuclei more accurately
~\cite{becvar,chyzh,kroll}. Only the DICEBOX gamma cascade for neutron capture on $^{157}$Gd was incorporated into the RAT-PAC-based simulation for this work, since the high thermal neutron capture cross section (254,000~b)~\cite{hagiwara} dominates the other isotopes. The associated parameters used to define the resonance structures of the $^{158}$Gd have been measured by the DANCE (Detector for Advanced Neutron Capture Experiments) collaboration and are available in the literature~\cite{chyzh}. 

The RAT-PAC-DICEBOX simulation was used to determine the WATCHBOY detector response for neutron capture throughout the target volume. For tuning purposes, the response was compared against the $^{252}$Cf calibration data taken in October 2013 and March 2014 (around the time period of physics data taking) and the neutrons resulting from muon spallation in the water target (during the time period of physics data taking). The latter was identified as events that pass the photoelectron and charge balance cuts and were within [20,200]~$\mu s$ since the last muon event~\cite{dazeley}. The water absorption length and the target wall reflectivity in the simulation were tuned to maximize agreement with the summed photoelectron and charge balance distributions obtained from the $^{252}$Cf calibration data and the neutron data resulting from muon spallation in the water target. The values of the absorption length and reflectivity are negatively correlated, so constraints on these two values must be analyzed together. To select optimal values and uncertainties for each, the data and RAT-PAC-DICEBOX simulations were compared 
over a range of input reflectivity and water absorption length values using a $\chi^2$ analysis (see Fig.~\ref{Fig:chi} and Fig.~\ref{Fig:chiMuonNeu}), revealing bands of good fits 
extending from the top left down to the right in both the photoelectron and charge balance plots. 
Example comparison between the simulation and the calibration data is shown in Fig.~\ref{Fig:cf}, and an example comparison between the simulation and the neutron data from muon spallation in water target is shown in Fig.~\ref{Fig:muonNeu}. In Fig.~\ref{Fig:cf}, wall reflectivity of 98\% and water attenuation length of 28~m that yielded $\chi^2$/ndf value of 1.6
for the photoelectron response and 2.1 for the charge balance  response were chosen. 
The poor $\chi^2$/ndf value for the charge balance response could potentially be explained by the target reflective coating that slightly bent, which introduced a disturbance in the light reflection and detection. There could also be variations in the PMT efficiency near the target wall that contributed to the variations in the PMT hit multiplicity.
The multiplicity and energy distributions of the gamma rays produced by the de-excitation of $^{156}$Gd nuclei that were not corrected for with DICEBOX could also affect the charge balance response especially in the case of neutron captures close to the target wall where gamma rays are more likely to escape. 
Additional subtle detector effects may also have contributed, such as PMT flashers events that somehow passed the charge balance cut, or other electronic noise. The effect of PMT dark rate on the data distributions was checked assuming a 10~kHz dark rate per PMT, which is relatively high, and found to be negligible. 
In Fig.~\ref{Fig:muonNeu}, a wall reflectivity of 90\% and a water attenuation length of 45~m that yielded $\chi^2$/ndf values of 1.4 and 1.2 for the photoelectron and charge balance responses were chosen, respectively.
Fig.~\ref{Fig:chi}~and~\ref{Fig:chiMuonNeu}  contains more than 700 comparisons between independent simulated data sets and the real data set containing 6$\times$10$^{5}$ and 2$\times$10$^{3}$ events, respectively. 

\begin{figure}[!ht]
\centering
\begin{minipage}{0.5\textwidth}
\includegraphics[width=\linewidth]{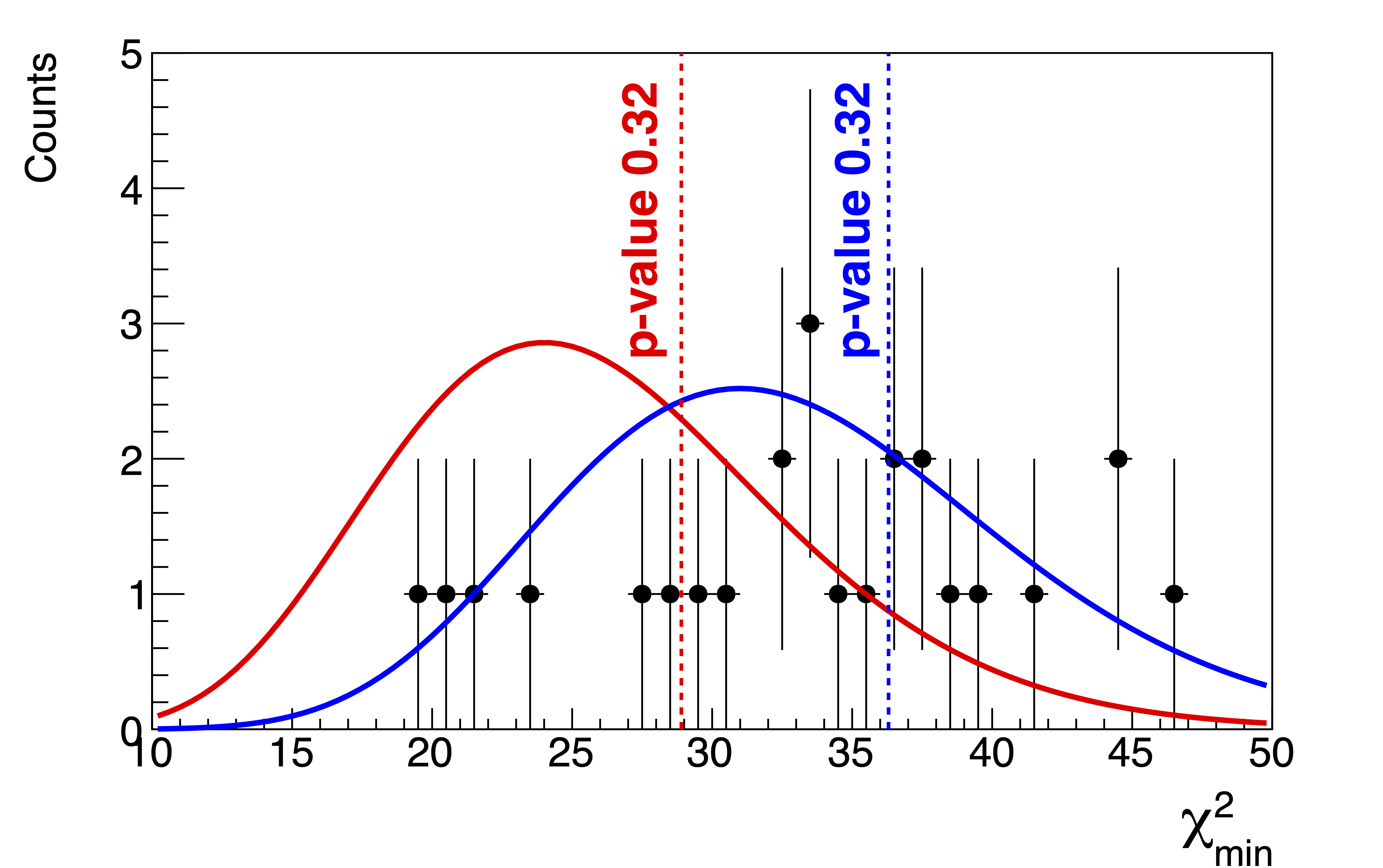}
\vspace{-0.4cm}
\end{minipage}
\caption{Distribution of the minimum $\chi^2$ values. The $\chi^2$ distributions with 26~DOF (red) and 33~DOF (blue) are shown. In our analysis, we accepted data points that fall within one standard deviation of the mean of $\chi^2$ distribution with 33~DOF (dashed blue).}\label{Fig:chiWithLim}
\end{figure}

The uncertainty in the neutron efficiency that results from the tuning process was determined by propagating the uncertainty in the wall reflectivity and absorption length through the analysis. To determine the uncertainties, we performed a series of 1D scans of $\chi^2$ values as a function of reflectivity for a series of absorption values spanning Fig.~\ref{Fig:chi} and Fig.~\ref{Fig:chiMuonNeu}. Each was fitted with a second degree parabolic function to determine the $\chi^2$ minimum. An example of a distribution of these $\chi^2$ minima for Fig.~\ref{Fig:chi}a is shown in Fig.~\ref{Fig:chiWithLim}. The degree of freedom (DOF) of the PE response is 26~DOF and the expected distribution of $\chi^2$ values for 26~DOF is shown in red (Fig.~\ref{Fig:chiWithLim}). While $\chi^2$ minima within 68\% confidence level (CL) for 26~DOF were observed (data points on the left side of dashed red line on Fig.~\ref{Fig:chiWithLim}), we noticed that the distribution of $\chi^2$ values for 33~DOF fitted our data best, indicating an imperfect match between simulation and data. The range of acceptable reflectivity and absorption length values were the ones that yielded a $\chi^2$ minimum value within 68\% CL for 33~DOF (data points on the left side of dashed blue line on Fig.~\ref{Fig:chiWithLim}). An identical method was performed for Fig.~\ref{Fig:chi}b, Fig.~\ref{Fig:chiMuonNeu}a,  and Fig.~\ref{Fig:chiMuonNeu}b to determine the acceptable range of wall reflectivities and absorption lengths. 
The accepted ranges are shown in red in Fig.~\ref{Fig:chi} and Fig.~\ref{Fig:chiMuonNeu}.
The detection efficiencies of neutrons uniformly distributed in the target volume were evaluated for the mean, upper bound and lower bound of each data point (red bars in Fig.~\ref{Fig:chi} and Fig.~\ref{Fig:chiMuonNeu}) to determine the range of uncertainties.
The detection efficiency of neutrons uniformly distributed in the target volume was found to be 30.3\% $\pm$ 5.3\%~(sys.).

\section{\label{sec:level4}Measurement of the detection rates of correlated neutron events}

In Section~\ref{sec:leve} we described the RAT-PAC-DICEBOX-GEANT4 simulation used to simulate the response of WATCHBOY to neutron capture events in the target. This simulation was used to determine the neutron detection efficiency. Independent and separate simulations were also done to determine the rate of neutron captures inside the WATCHBOY target. The simulations were performed with the input neutron energy spectrum measurement of MARS, which was incident upon the outer WATCHBOY shield, which also acts as a muon detector. Two such independent simulations were performed. One was based on RAT-PAC-GEANT4, the other was based on FLUKA.

\begin{figure}[!ht]
\centering
\begin{minipage}{0.50\textwidth}
\includegraphics[width=\linewidth]{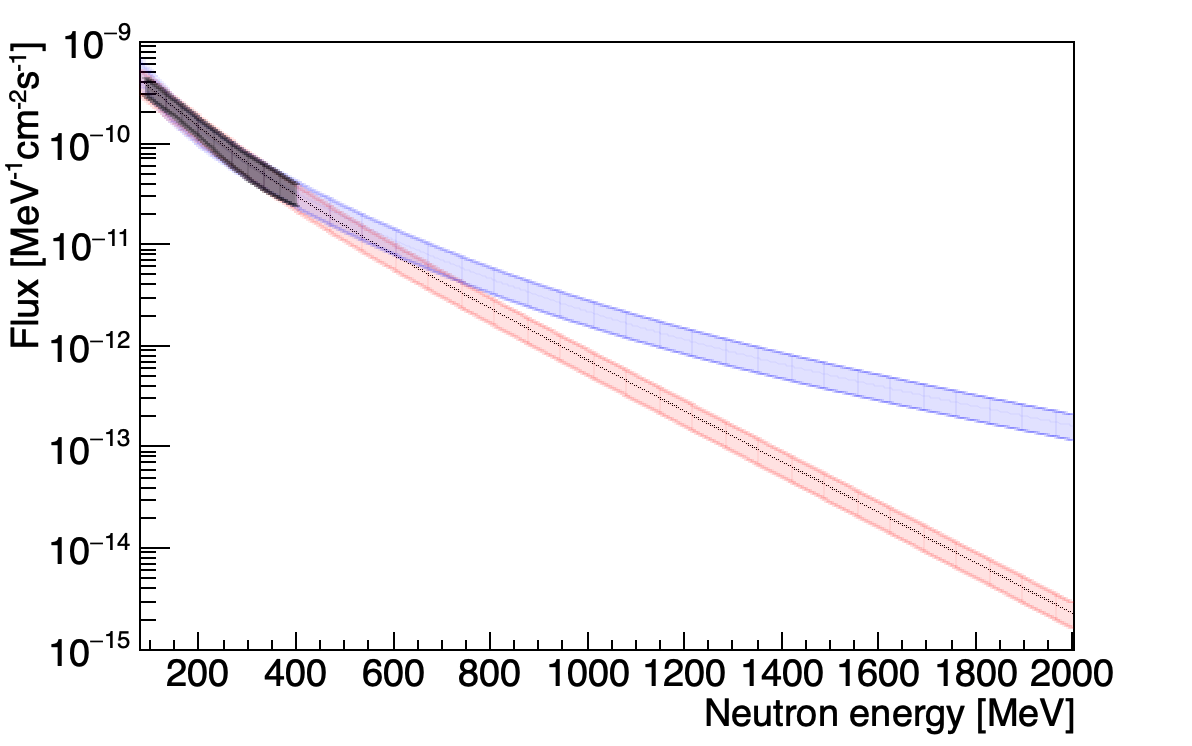}
\vspace{-0.4cm}
\end{minipage}
\caption{(Color online) The MARS neutron energy spectrum (black shaded region) propagated over a broad range of energies (from 20 MeV to 2 GeV), assuming two different scaling models; a double exponential (red), and Mei and Hime (blue). The flux uncertainty of the MARS data was propagated over the whole energy range for both models.}
\label{Fig:marsData}
\end{figure}

The MARS fast neutron spectrum measured at the same depth as the WATCHBOY detector extends from 90~MeV to 400~MeV (Fig.~\ref{Fig:marsData}, black). No information was available for how the neutron energies scale above 400 MeV at this depth. However we explored the effect of two different scaling models above 400 MeV - a double exponential function (red), chosen to approximate the same shape at higher energies, and the Mei and Hime model~\cite{mei} (blue). The two are shown in Fig.~\ref{Fig:marsData}. The differences in the resulting neutron detection rates were evaluated and used to determine the systematic uncertainty due to the unmeasured neutron spectrum above 400 MeV. The two different scaling assumptions begin to diverge strongly above $\approx$700~MeV. We note that according to the simulation using the harder Mei and Hime energy spectrum, only 2\% of the neutron captures in the WATCHBOY target result from those above 700~MeV, so the uncertainty in the shape above this energy contributes a relatively small amount to the total uncertainty.

The angular distribution of the neutron flux at 390 ($\pm$ 12~m.w.e) incident upon the WATCHBOY detector was unknown. However, it has been measured above ground and found to be consistent with $\approx$cos$^3 \theta$~\cite{Ziegler}. Below ground, we expect the angular distribution to be less peaked along the muon direction than above ground - an energy-dependent combination of isotropic and peaked along the initiating muon direction~\cite{mei}. The neutron initial directions were therefore simulated at cos$\theta^3$, cos$\theta^2$ and cos$\theta$, and the differences between the results of these simulations were used to estimate our systematic uncertainty due to uncertainties in the angular distribution. 

\begin{figure}[!ht]
\centering
\begin{minipage}{0.5\textwidth}
\includegraphics[width=\linewidth]{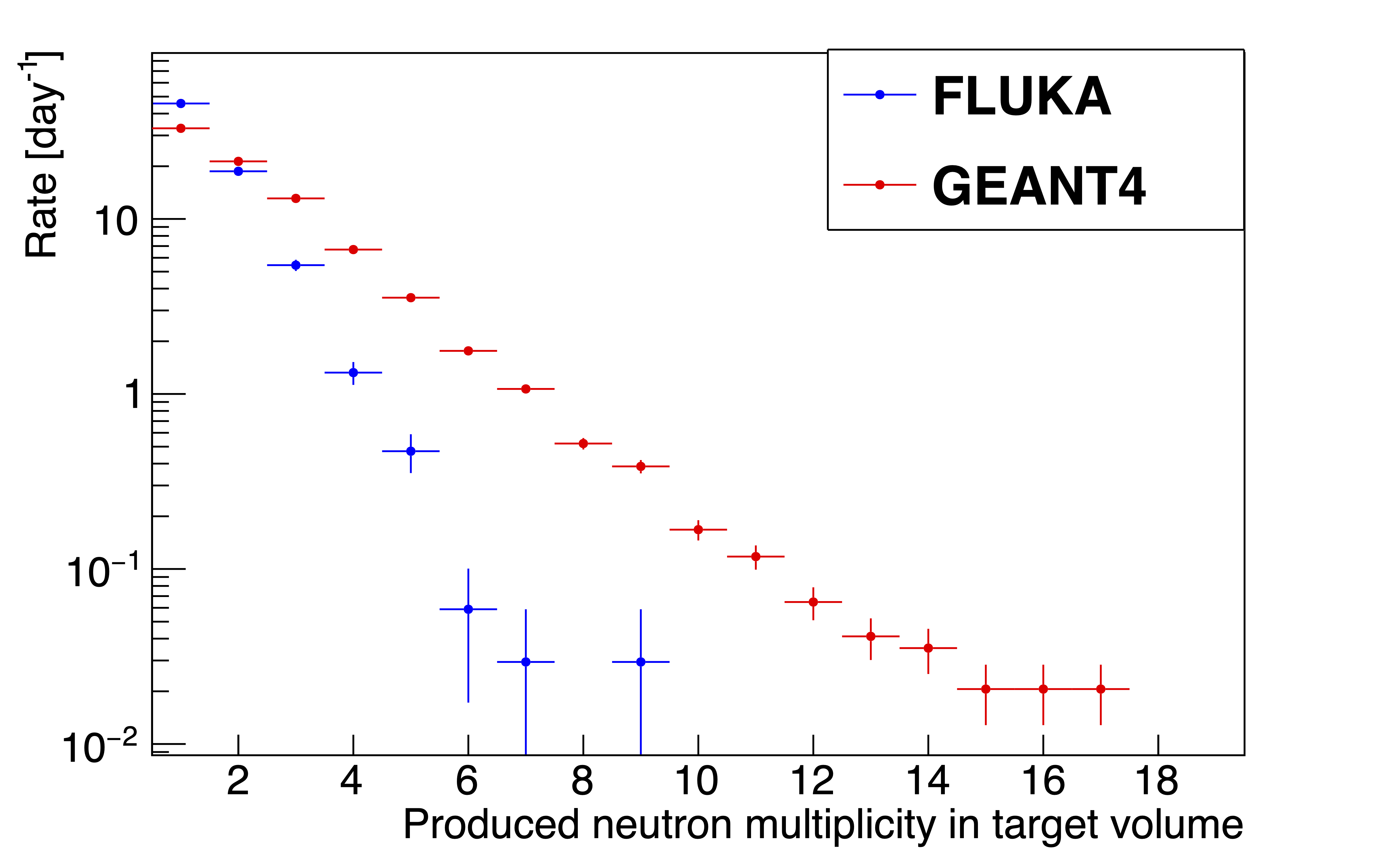}
\vspace{-0.4cm}
\end{minipage}
\caption{(Color online) The distributions of the rate of produced event clusters in the target volume obtained from the fast neutron simulations in FLUKA and GEANT4.}\label{Fig:daughter}
\end{figure}

\begin{figure}[!ht]
\centering
\begin{minipage}{0.5\textwidth}
\includegraphics[width=\linewidth]{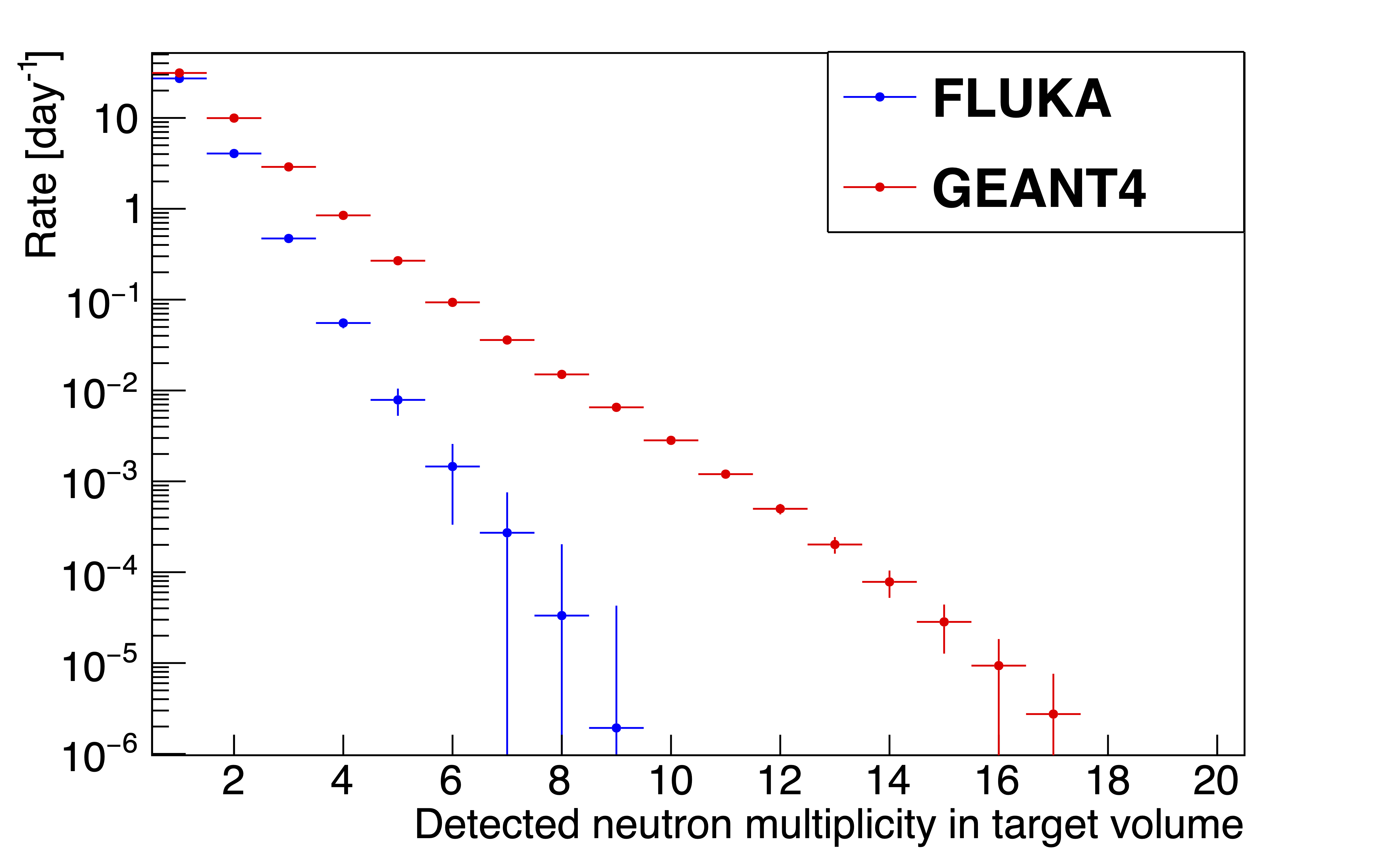}
\vspace{-0.4cm}
\end{minipage}
\caption{(Color online) The distributions of detected event clusters that pass the analysis cuts obtained from the fast neutron simulations in FLUKA and GEANT4. The GEANT4-based distribution has a higher average multiplicity in comparison with the FLUKA-based result.}\label{Fig:grandDau}
\end{figure}

The RAT-PAC-GEANT4 and FLUKA-based distributions of neutron multiplicity in target volume (Fig.~\ref{Fig:daughter}) were obtained from the above simulations. They were then converted to detected neutron multiplicity distributions (Fig.~\ref{Fig:grandDau}) using the following binomial equation:
\begin{equation} \label{Eq:grandDau}
M_i= \sum_{k=i}^{\infty} {k\choose i} \epsilon^{i} (1-\epsilon)^{k-i} P_k 
\end{equation}

where $M_i$ is the probability of the detected neutron multiplicity equals to i, $\epsilon$ is the thermal neutron detection efficiency, and $P_k$ is the probability of multiplicity of secondary neutrons created per primary equals to k. 

The resulting RAT-PAC-GEANT4-based distribution of neutron multiplicity yielded a flatter distribution and higher average (2.3~neutrons) than the FLUKA-based distribution (1.5~neutrons). Likewise, the RAT-PAC-GEANT4-based detected neutron multiplicity distribution was also flatter and had a higher average (1.45~neutrons) than the FLUKA-based one (1.17~neutrons).

\begin{figure}[!ht]
\centering
\begin{minipage}{0.50\textwidth}
\includegraphics[width=\linewidth]{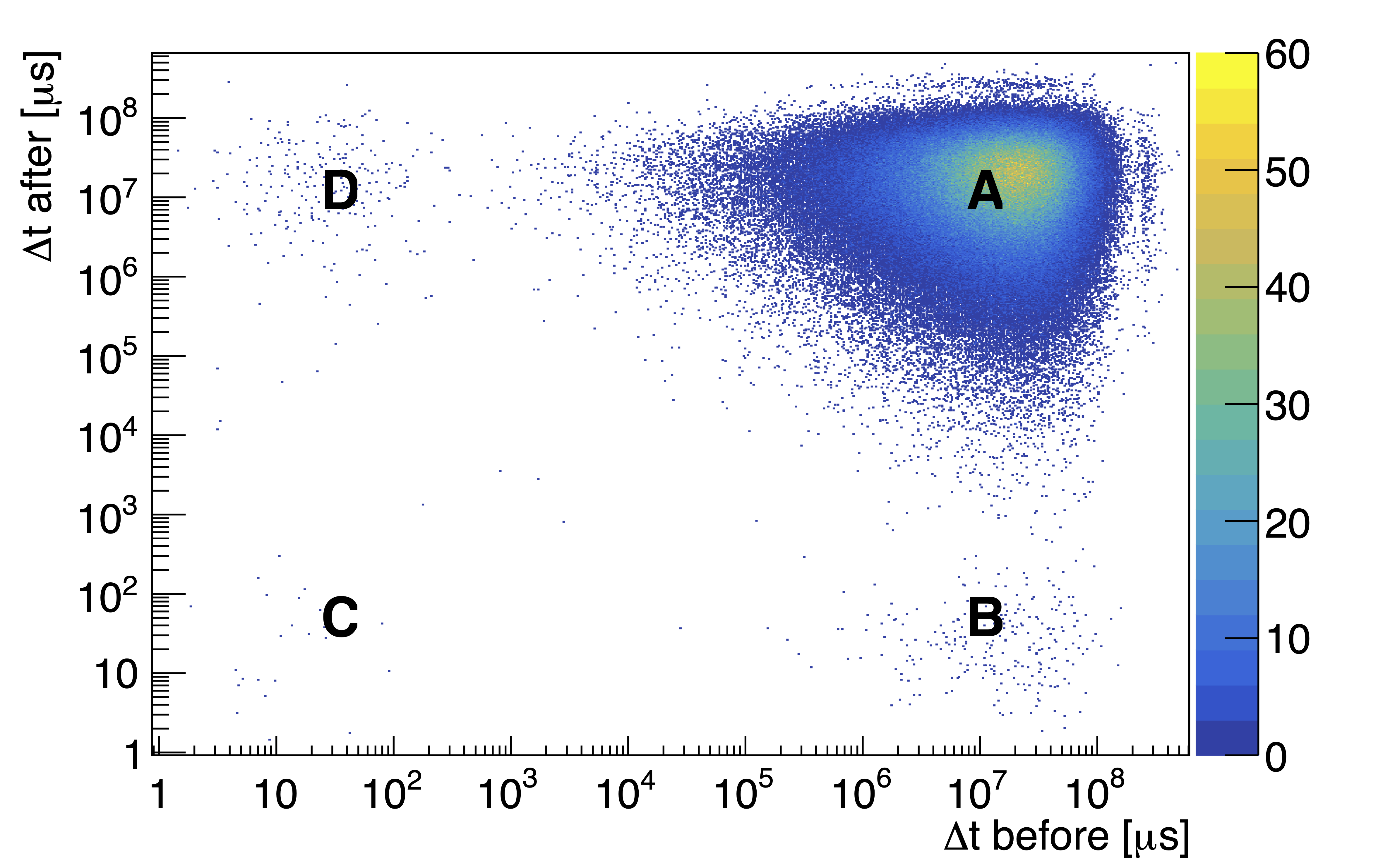}
\vspace{-0.4cm}
\end{minipage}
\caption{The measured distribution of time intervals between detected physics events. Region B, C, and D likely correspond to the correlated neutron events and region A likely corresponds to single events.}\label{Fig:phy_t2mu}
\end{figure}

To calculate the degree to which accidental coincidences of single events contribute to the rate of detected multiple coincidence event clusters, a simple toy Monte Carlo (toy MC) simulation was performed ($10^8$ events). Note, in the following we will use the term ``event cluster'' to refer to a collection of correlated events. The toy MC provides a series of timestamps, each corresponding to either a detected single event cluster (which always has a multiplicity equals to one) or a detected neutron event cluster. To perform this simulation, the detected single rate and the detected neutron multiplicity distribution, presented in Fig.~\ref{Fig:grandDau}, must be known. The detected single rate was approximated based on the observed events that dominated region A in Fig.~\ref{Fig:phy_t2mu}. 
The toy MC was performed as follows:

\begin{itemize}[noitemsep]
    \item The time to the next event cluster is sampled as $dt= -log(\xi)/R_{total}$, where $\xi$ is a random number from 0 to 1 and  $R_{total}$ is the sum of the detected single rate ($R_s$) and the detected neutron cluster rate. Note, the detected neutron cluster rate is the detected neutron rate divided by the average multiplicity of the detected neutron events ($R_{mul}$).
    \item Based on $R_s$ and $R_{mul}$, the type of the next event cluster is determined.
    \item If the event cluster type is a detected single event,  we proceed to the next event cluster.
    \item If the event cluster type is a detected neutron event cluster, the multiplicity is sampled from the distribution shown in  Fig.~\ref{Fig:grandDau}. The time of each detected neutron capture within each cluster is sampled as $dt= -\log(\xi)/26\mu s$ relative to the initial time, where 26$~\mu s$ is the experimentally determined average inter-event time between neutron captures in Region B, C and D from  Fig.~\ref{Fig:t2mu} and Fig.~\ref{Fig:phy_t2mu}.
\end{itemize}

\begin{figure}[!ht]
\centering
\begin{minipage}{0.5\textwidth}
\includegraphics[width=\linewidth]{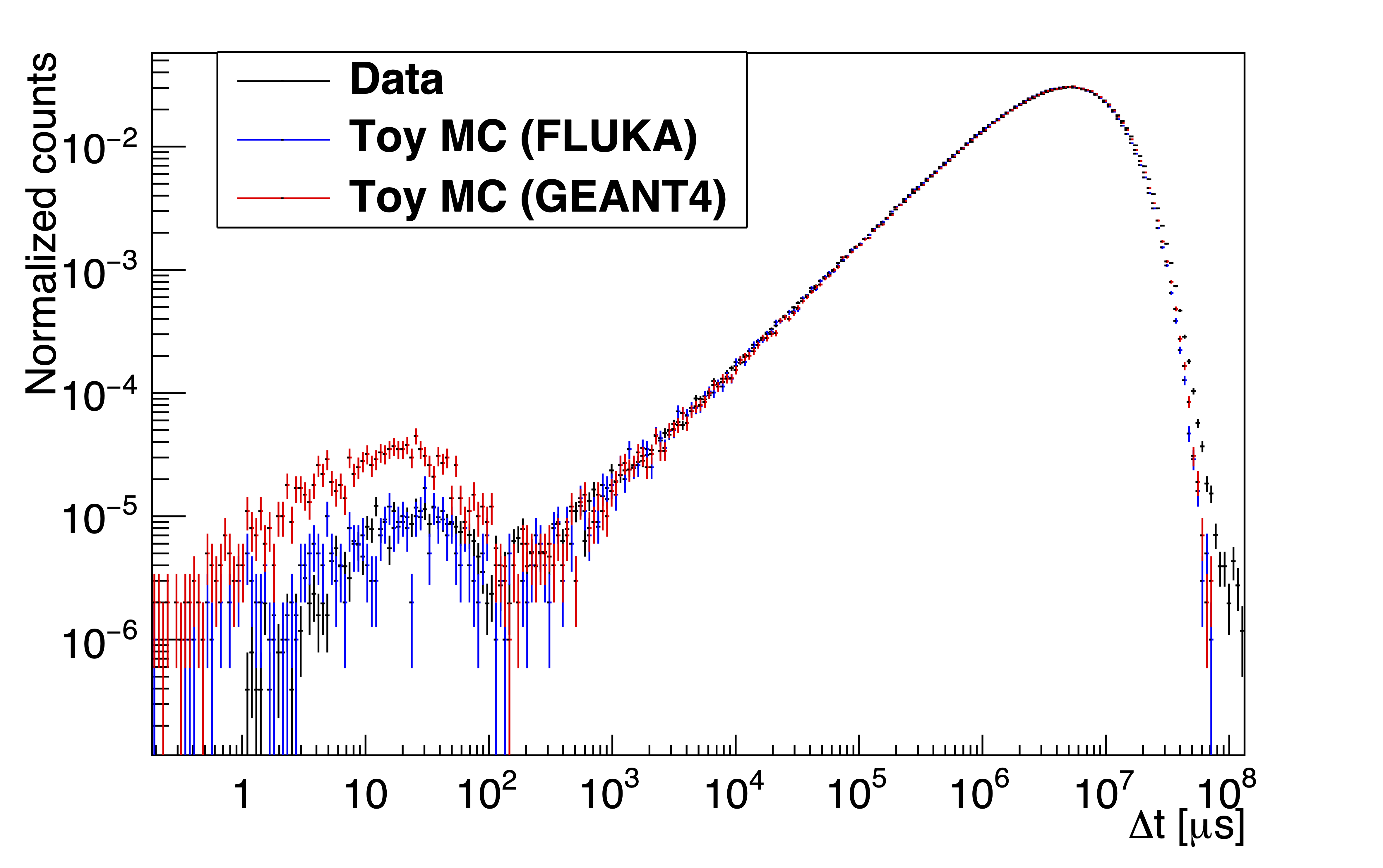}
\vspace{-0.4cm}
\end{minipage}
\caption{(Color online) The X-projection of the distribution of time intervals between the detected physics events is compared against the simulated distribution of the inter-event time obtained from a toy Monte Carlo simulation.}\label{Fig:mc}
\end{figure}

The toy MC-based distributions of inter-event times could then be realized and compared with the data as shown in Fig.~\ref{Fig:mc}. The data inter-event time distribution was generated from the X-projection of Fig.~\ref{Fig:phy_t2mu}. 
The $\chi^{2}$/NDF comparison between data and the two toy MC distributions between $10^{-1}$~$\mu s$ and $10^{7}$~$\mu s$ yielded values of 1.84 and 3.95 for FLUKA and GEANT4 respectively. Inter-event times greater than $10^{7}$~$\mu s$ were excluded in the $\chi^{2}$ analysis because of the appearance of artifacts in the data that were noted at long inter-event times due to DAQ resets that occurred when the digitizer buffers were read out. 

\begin{figure}[!ht]
\centering
\begin{minipage}{0.5\textwidth}
\includegraphics[width=\linewidth]{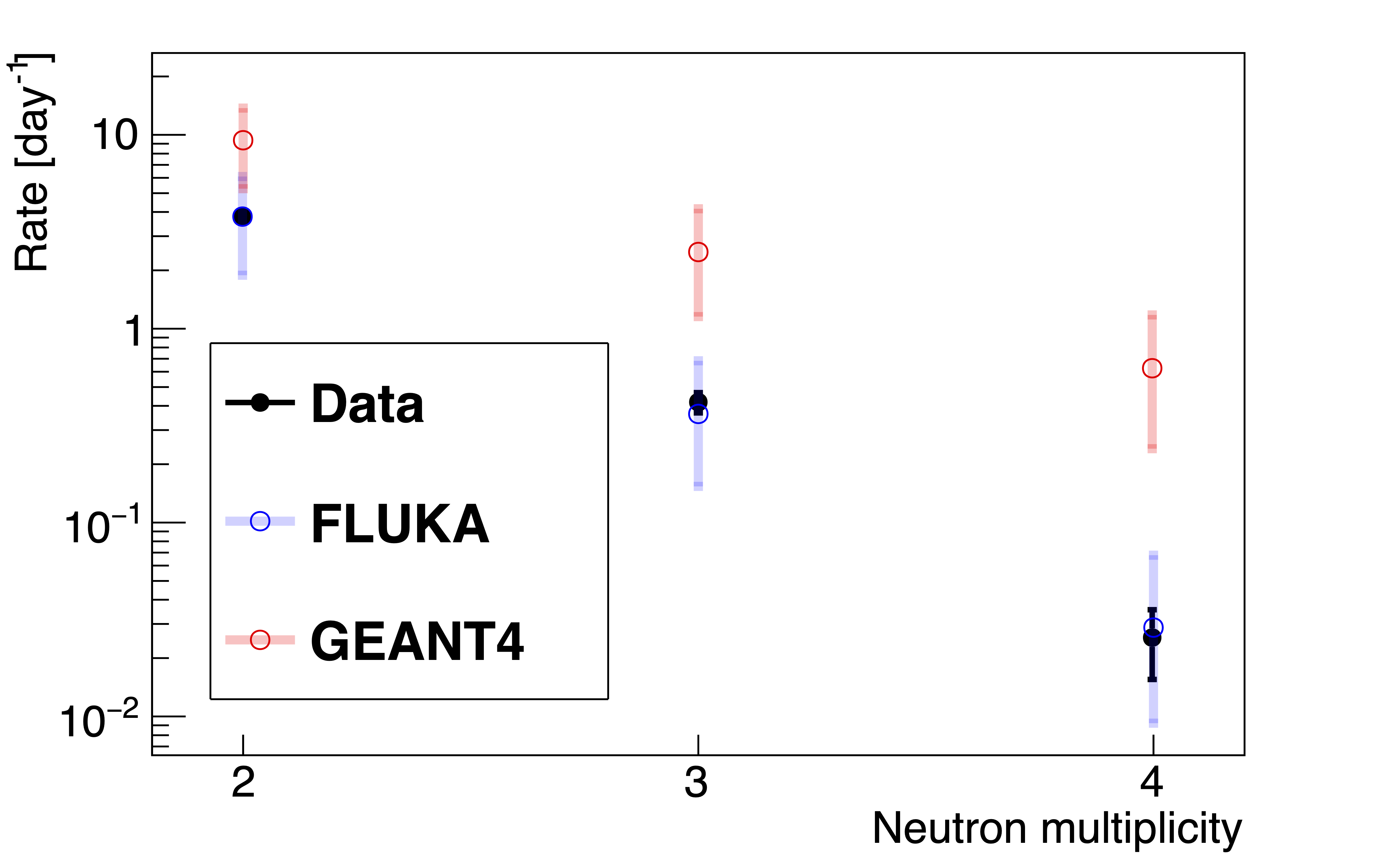}
\vspace{-0.4cm}
\end{minipage}
\caption{(Color online) The distribution of the simulated and measured neutron-like multiplicity after contamination from single events is incorporated.  The GEANT4-based detection rate of correlated pairs was approximately 2 times higher than the FLUKA-based one. A good agreement between the measured and the FLUKA-based detected neutron-like event multiplicity rate was observed.}\label{Fig:mul}
\end{figure}

A 100~$\mu s$ time cut (red line on Fig.~\ref{Fig:t2mu}) was applied to both the toy MC and data to obtain the neutron multiplicity distributions shown in Fig.~\ref{Fig:mul}. While good agreement was observed between the measured data and FLUKA-based distribution, poor agreement was observed for the GEANT4-based result. This suggests that the Bertini cascade model used in the standard Shielding physics list in the RAT-PAC-GEANT4 simulation overestimated the neutron multiplicities in water-based media.

Table.~\ref{tab:parameter} shows the systematic uncertainties 
that result from the uncertainties in the simulation input parameters. The positive and negative systematic uncertainties for each simulation parameter were estimated by varying the simulation parameter of interest while keeping the other simulation parameters constant. For the fast neutron spectrum and the neutron efficiency parameters, the positive (negative) systematic uncertainty was the difference between the highest (lowest) detection rate and the median detection rate. For the cosine parameter, we cannot assume that the distribution is asymmetric and hence the positive (negative) systematic uncertainty was the difference between the highest (lowest) detection rate and the mean detection rate.

\begin{table}[!t]
\caption{The relative systematic uncertainty associated with variations of the simulation parameters}
\label{tab:parameter}
\centering
\begin{tabular}{|c || c c | }
\hline
Simulation input &          FLUKA        &      GEANT4     \\
\hline
DINEUTRON RATE         &                 &                 \\
Cosine                 &1.3\%            &1.2\%            \\
Fast neutron spectrum  &+22.7\% -30.4\%  &+21.7\% -27.7\%   \\
Neutron efficiency     &+31.7\% -28.6\%  &+20.4\% -21.6\%    \\
\hline
TRINEUTRON RATE        &                 &                 \\
Cosine                 &1.9\%            &1.1\%            \\
Fast neutron spectrum  &+27.4\% -36.3\%  &+22.4\% -29.9\%   \\
Neutron efficiency     &+53.0\% -40.3\%  &+36.6\% -32.9\%    \\
\hline
QUADNEUTRON RATE       &                 &                 \\
Cosine                 &3.4\%            &0.95\%           \\
Fast neutron spectrum  &+38.0\% -41.8\%  &+22.2\% -32.9\%  \\
Neutron efficiency     &+71.3\% -47.7\%  &+54.7\%  -45.8\%  \\
\hline
\end{tabular}
\end{table}

Other potential backgrounds considered were decays due to muon--induced radionuclides (e.g. $^9$Li/$^8$He), and antineutrinos from world reactors. Both of these can produce correlated neutron-like events. A measurement of the rate of  muon-induced $^9$Li in WATCHBOY was attempted previously and an upper limit was published~\cite{dazeley}. The published uncertainty in the di-neutron-like event rate in WATCHBOY was negligible (2.7\%) relative to the other uncertainties investigated here and was not included in our analysis. The nearest reactor (McGuire Nuclear Station, 2$\times$3411~MW, 216~km away) is too distant to produce a significant rate of antineutrino interactions in the 2 ton WATCHBOY target. According to \cite{barna}, the rate of IBD events due to the integrated world reactor flux, plus geo-antineutrinos in WATCHBOY is only 0.0014 events per day, which produces a negligible contribution to the correlated neutron signal.

\begin {comment}
Previously, we estimated the upper limit of $^9$Li production rate to be $1.9 \times 10^{-7} \mu^{-1} g^{-1} cm^2$~\cite{dazeley}. Here we used a different approach to re-estimate the upper limit of $^9$Li production rate.
The measured and the FLUKA-based detection rates of two-correlated neutrons were found to be 3.77 $\pm$ 0.2 (stats.) and 4.5 $\pm$ 0.03 (stats.) + 1.5 (sys.) - 1.2 (sys.), respectively (Fig.\ref{Fig:mul}). These uncertainties corresponded to 1-sigma or 68\% CL. 
Using 1.645~sigma or 90\% CL, we estimated the FLUKA-based two-correlated neutron detection rate to be  4.5 $\pm$ 0.03 (stats.) + 2.5 (sys.) - 2.0 (sys.). 
Eq.(\ref{eq:estimateLi}) can be used to estimate the upper limit on the background contribution due to muon-induced $^9$Li~\cite{dazeley}.

\begin{equation} \label{eq:estimateLi}
    Y \textless \frac{N}{\epsilon B_{\beta n} T L_{\mu} R_{\mu} \rho} = \frac{R}{L_{\mu} R_{\mu} \rho}
\end{equation}

N is the upper limit on the number of observed $^9$Li candidates, $\epsilon$ is the estimated detection efficiency (25.2$\pm$5.4\%)~\cite{dazeley}, $B_{\beta n}$ is the $^9$Li $\beta$-neutron branching ratio (50.8$\pm$0.9\%)~\cite{dazeley}, T is the livetime, $L_{\mu}$ the average muon path length through the target (87.0 $\pm$ 3.1~cm)~\cite{dazeley}, $R_{\mu}$ the estimated muon rate (0.9 $\pm$ 0.007~Hz)~\cite{dazeley}, $\rho$ the target density(1.0~$g/cm^3$), and $R$ is the excess of the detection rate of two-correlated neutrons ( 3.77 - (4.5-2.0) )/day = 1.27/day. 
We estimated the upper limit of the production rate of $^9$Li in water to be $1.9 \times 10^{−7} \mu^{-1} g^{-1} cm^2$, which confirmed our previous estimate~\cite{dazeley} and the upper limit set by Super Kamiokande measurement ($0.5 \times 10^{−7} \mu^{-1} g^{-1} cm^2$)~\cite{zhang}.

\end{comment}

\section{\label{sec:level5}Conclusions}

The rate of correlated neutron-like event clusters in the WATCHBOY target was measured and found to be consistent with the expected rate from a FLUKA simulation of the MARS fast neutron spectrum incident upon the WATCHBOY outer water shield. 
The observed agreement provides an independent confirmation of the MARS fast neutron spectrum measured at the same depth as WATCHBOY (390~m.w.e.). Which in turn increases confidence in the MARS fast neutron spectra measured at other overburdens. A similar agreement was not found with a GEANT4-based simulation. Since MARS was designed to be transportable, and was deployed at multiple depths in the same configuration, the systematic bias between its measurements was minimized. Hence, a predictive model of the high-energy neutron energy-dependent flux can be obtained at any depth~\cite{roecker}. Confirmation of the MARS fast neutron spectrum places an important constraint on the limited number of depth scaling models that are used to extrapolate fast neutron flux between different overburdens and to validate the simulation framework that would be used to predict the rate of correlated neutrons in large underground water-based detectors planned for future construction, such as the WATCHMAN detector. 

\section{\label{sec:levelx}Acknowledgement}

The authors would like to thank S. Valenta for providing the DICEBOX model, J. Xu for the helpful discussion on $\chi^2$ analysis, and V. Kudryavtsev for helpful discussion on the fast neutron simulation. The authors also acknowledge AIT-WATCHMAN collaboration, M. Sweany, P. Marleau, B. Cabrera-Palmer, M. Gerling, K. Vetter, and T.M. Shokair for contributing to the WATCHBOY and MARS detector design, implementation, simulation, and characterization. Part of this work was performed under the auspices of the US Department of Energy by Lawrence Livermore National Laboratory under contract DE--AC52--07NA27344, release number LLNL-JRNL-796592. The research of F.S. was performed under the appointment to the Lawrence Livermore Graduate Scholar Program Fellowship and Rackham Merit Fellowship. The work of I.J. and F.S. was partially supported by the Consortium for Verification Technology under U.S. Department of Energy National Nuclear Security Administration award number DE--NA0002534.

\section{\label{sec:levelxx}Appendix A: Estimation of rate of detectable events due to PMT radioactivity}

We also validated our simulation toolkit by using it to predict the rate of detectable events due to PMT radioactivity and then compare the results with the observation.

\begin{figure}[!ht]
\centering
\begin{minipage}{0.50\textwidth}
\includegraphics[width=\linewidth]{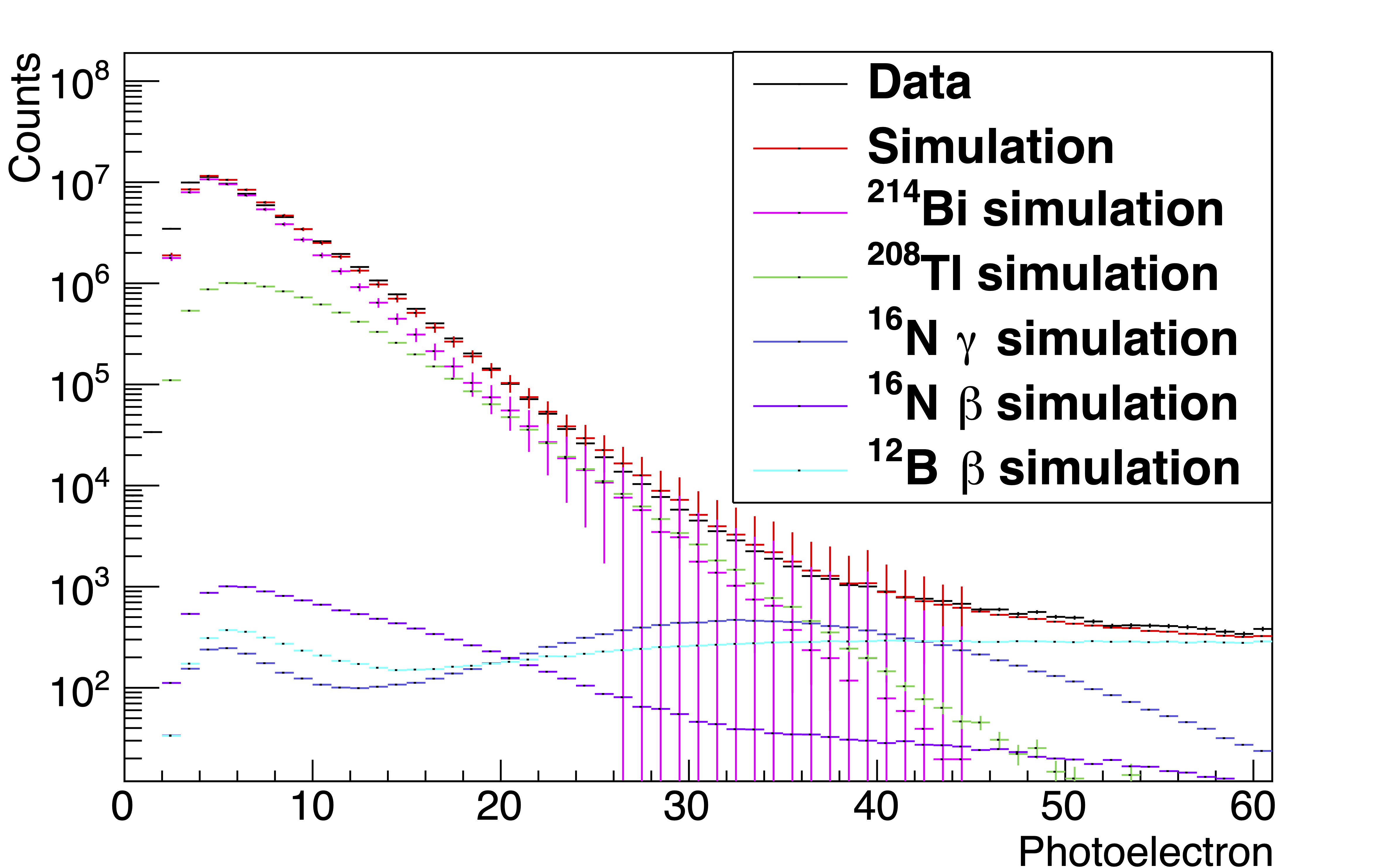}
\vspace{-0.4cm}
\end{minipage}
\caption{The simulated detector response to $^{214}$Bi is scaled and compared against the experimental PE response of single events.}\label{Fig:214bi}
\end{figure}

The detection rate of events due to PMT radioactivity can be approximated by simulating interactions of uranium progeny such as $^{214}$Bi in the WATCHBOY detector, obtaining its simulated detection efficiency, and scaling its simulated PE response to match the experimental PE response of single events as shown in Fig.~\ref{Fig:214bi}. The estimated rate is $\approx$7.2/PMT/s, which is in decent agreement with the average rate for Hamamatsu R7081 PMT ($\approx$5.9~Bq/PMT for the Uranium chain). 
 
\section{\label{sec:levelxx}References}

 
%

\end{document}